\documentclass[12pt,preprint]{aastex}

\usepackage{natbib}
\usepackage{graphicx}
\usepackage{amsmath}

\usepackage{ulem}

\renewcommand{\vec}{\boldsymbol}
\newcommand{\tens}{\boldsymbol}
\newcommand{\bm}{\overline{\vec{B}}}
\newcommand{\vm}{\overline{\vec{v}}}
\shorttitle{Dipole Collapse and Dynamo Waves}
\shortauthors{Schrinner et al.}

\begin{document}
\title{Dipole Collapse and Dynamo Waves\\ in Global Direct Numerical 
Simulations}

\author{Martin Schrinner, Ludovic Petitdemange\altaffilmark{1}
\altaffiltext{1}{Previously at
    Max-Planck-Institut f\"{u}r Astronomie, K\"{o}nigstuhl 17, 69117
    Heidelberg, Germany} \and  Emmanuel Dormy}
\affil{MAG(ENS/IPGP), LRA, \'Ecole Normale Sup\'erieure, 24 Rue Lhomond, 
  75252 Paris Cedex 05, France}
\email{martin@schrinner.eu}

\begin{abstract}
Magnetic fields of low-mass stars and planets are thought to 
originate from self-excited dynamo action in their convective 
interiors. Observations reveal a variety of field topologies ranging 
from large-scale, axial dipole to more structured magnetic fields. In this 
article, we investigate more than 70 three-dimensional, self-consistent 
dynamo models obtained by direct numerical simulations. The control 
parameters, the aspect ratio and the mechanical boundary conditions 
have been varied to build up this sample of models. Both, strongly 
dipolar and multipolar models have been obtained. We show that these dynamo 
regimes can in general be distinguished by the ratio of a typical convective 
length-scale to the Rossby radius. Models with a predominantly dipolar 
magnetic field were obtained, if the convective length scale is at least an 
order of magnitude larger than the Rossby radius. Moreover, we highlight 
the role of the strong shear associated with the geostrophic zonal 
flow for models with stress-free boundary conditions. In this case the 
above transition disappears and is replaced by a region of bistability for 
which dipolar and multipolar dynamos co-exist. We interpret our results 
in terms of dynamo eigenmodes using the so-called test field method. We can 
thus show that models in the dipolar regime are characterized by an 
isolated `single mode'. Competing overtones become significant as the 
boundary to multipolar dynamos is approached. We discuss how these 
findings relate to previous models and to observations.
\end{abstract}
\section{Introduction}
Topologies of stellar and planetary magnetic fields stemming from 
hydrodynamic dynamo action are highly variable. Observations revealed 
complicated field configurations dominated by higher multipoles, but also
large-scale or even dipolar stellar and planetary magnetic fields 
are ubiquitous \citep{anderson10,hulot10,russell10,donati09}. The latter
is the more astonishing, as the magnetic Reynolds number in planetary 
and stellar convection zones is often large and most hydrodynamic 
dynamos operate far above their threshold.  
Observations and numerical simulations indicate that rapid global rotation and 
thus the ordering influence of the Coriolis force is of major importance for 
the generation of coherent magnetic fields \citep{stellmach04,
kapylae09,brown10}. \cite{kutzner02} demonstrated the existence of a 
dipolar and a multipolar dynamo regime and \cite{christensen06} showed that
the transition between both is governed by a local Rossby number, 
i.e. the influence of inertia relative to the Coriolis force. Similar results 
were reported by \cite{sreenivasan06}, too. Dipolar models were found for 
small Rossby numbers; they are separated by a fairly sharp regime boundary 
from multipolar models, for which inertia is more important.

\cite{sreenivasan11} argued that dipolar magnetic fields enhance the 
kinetic helicity and are therefore easier to maintain than fields with a more 
complicated field topology. However, as noted by \cite{sreenivasan11}, the
relation between kinetic helicity and induction mechanisms is 
not straightforward. Moreover, \cite{schrinner07} showed that the 
kinetic helicity is indeed a bad proxy measure for the induction effects 
(\(\alpha\)-effect) in these models.
    
\cite{schrinner10a} pointed out that the high relative dipole field 
strength for models in the low Rossby number regime is associated with 
the dominance of only one dipolar eigenmode. Spatially more complex modes 
were found to be strongly diffusive. In this dynamo regime, higher order 
contributions to the magnetic field result mainly from the deformation of the 
fundamental mode by the turbulent flow. Subsequently, the small scale 
contributions decay due to ohmic diffusion \citep{hoyng09,schrinner11c}. As a 
result of the dominance of only one, real eigenmode, the axis of the dipole 
field is stable, polarity reversals or oscillations of the magnetic field do 
not occur.  

However, the reason for the dominance of an isolated eigenmode at low 
Rossby numbers and the cause of the dipole breakdown with decreasing 
influence of the Coriolis force are at present not well understood. 
In this study, we investigate how rotation influences the field 
topology in dynamo models in some more detail. We show that 
rotation-dominated convection in a spherical shell leads 
to a distinctive azimuthal field morphology which is well represented by the 
fundamental dynamo mode, but very different from the field pattern of the 
following higher-order modes. We argue that this discrepancy is 
responsible for the clear preference of the fundamental mode over 
higher-order modes. Furthermore, we discuss how the dominance of the 
fundamental mode is reduced if convection becomes less affected by rotation 
and the typical length scale of convection decreases relative to the Rossby 
radius. Because the topology and the time variability of the magnetic field in 
numerical dynamo simulations are closely correlated we likewise address the 
question under which conditions models exhibit fairly coherent oscillations, 
irregular polarity reversals or a stable dipole field.

Our analysis is based on 72 dynamo models in the Boussinesq approximation 
with different aspect ratios and mechanical boundary conditions. We generalize 
here the Rossby number criterion given by \cite{christensen06} to our sample 
of models. For some of them, we compute mean-field coefficients with the 
help of the test-field method \citep{schrinner07} to reveal their dynamo 
mechanisms.

\section{Dynamo Calculations}
Our dynamo models are solutions of the MHD-equations for a conducting 
Boussinesq fluid in a rotating spherical shell. The fluid motion is driven by
convection due to an imposed temperature difference, \(\Delta T\), between 
the inner and the outer shell boundaries. The fundamental length scale
of our models is the shell width \(L\), we scale time by \(L^2/\nu\), with 
\(\nu\) the kinematic viscosity, and temperature is scaled by \(\Delta T\). 
Moreover, following \cite{christensen01}, the magnetic field is considered in 
units of \(\sqrt{\varrho\mu\eta\Omega}\), with \(\varrho\) denoting the 
density, \(\mu\) the magnetic permeability, \(\eta\) the magnetic diffusivity 
and \(\Omega\) the rotation rate. With these units, the dimensionless 
momentum, temperature and induction equation are

\begin{eqnarray}
E\left(\frac{\partial\vec{v}}{\partial t}+\vec{v}\cdot\nabla\vec{v}-\nabla^2\vec{v}\right)
+2\vec{z}\times\vec{v}+\nabla P =\nonumber\\
Ra\frac{\boldsymbol{r}}{r_o}T
+\frac{1}{Pm}(\nabla\times\vec{B})\times\vec{B}\, ,\label{eq:dc:2}\\
\frac{\partial T}{\partial t}+\vec{v}\cdot\nabla T  =
\frac{1}{Pr}\nabla^2 T \, ,\label{eq:dc:4}\\
\frac{\partial\vec{B}}{\partial t}= \nabla\times(\vec{v}\times\vec{B})
+\frac{1}{Pm}\nabla^2\vec{B} \label{eq:dc:6}.
\end{eqnarray} 
We also note that \(\vec{v}\) and \(\vec{B}\) are solenoidal.
The system of equations is governed by four dimensionless parameters, the 
Ekman number \(E=\nu/\Omega L^2\), the (modified) Rayleigh number 
\(Ra=\alpha_T g_0\Delta T L/\nu\Omega\), the Prandtl number \(Pr=\nu/\kappa\), 
and the magnetic Prandtl number \(Pm=\nu/\eta\). In these definitions,  
\(\alpha_T\) stands for the thermal expansion coefficient, \(g_o\) is 
the gravitational acceleration at the outer boundary, and \(\kappa\) is the 
thermal diffusivity. A further control parameter is the aspect ratio of the 
shell defined as the ratio of the inner to the outer shell radius,  
\(\chi=r_i/r_o\). It determines the width of the convection zone. 

The mechanical boundary conditions are either i) no slip at the inner and the 
outer boundary, ii) no slip at the inner and stress free at the outer
boundary or iii) stress free at both boundaries. In the latter case, the 
angular momentum in the direction of the rotation axis was conserved. 
Furthermore, the magnetic field matches a potential field outside the fluid 
shell and fixed temperatures are prescribed at both boundaries.   

Some of the models investigated here exhibit bistability, i.e. the solution 
realized depends on the initial conditions for the magnetic field. We started 
most of the simulations with a dipolar magnetic field, but varied its initial 
amplitude over several orders of magnitude. Some calculations were started
from another model with slightly different parameters to test for hysteresis. 
In the bistable regime, models resulting from simulations with an initially 
weak magnetic field are here referred to as the `non-dipolar branch'. They are 
distinguished from dipolar solutions initially started with a strong magnetic 
field.
  
The numerical solver used to compute solutions of equations 
(\ref{eq:dc:2})-(\ref{eq:dc:6}) is PaRoDy (\cite{dormy98} and further 
developments). The numerical method is similar to that described 
in \cite{G84} except for the radial discretisation, which is treated in 
physical space on a stretched grid (allowing for a parallelization by a radial 
domain decomposition). Moreover, the pressure term has been eliminated by 
considering the double curl of the momentum equation.

\section{Non-Dimensional Output Parameters}
\label{out_param}
Our numerical dynamo-models are characterized by non-dimensional
output parameters. Dimensionless measures for the flow velocity are the 
magnetic Reynolds number, \(Rm=v_{\mathrm{rms}}\,L/\eta\), and the Rossby 
number, \(Ro=v_{\mathrm{rms}}/\Omega L\). In both definitions, 
\(v_{\mathrm{rms}}\) stands for the rms velocity of the flow. Similarly, 
\(B_{\mathrm{rms}}\) denotes the rms field strength.
We also introduce a local Rossby number, 
\(Ro_\ell=Ro\,\,\overline{\ell}_p/\pi\), based on the mean harmonic degree 
\(\overline{\ell}_p\) of the poloidal velocity field,
\begin{equation}
\overline{\ell}_p=\sum_\ell \ell\frac{<\boldsymbol{(v_p)}_\ell\cdot\boldsymbol{(v_p)}_\ell>}{<\boldsymbol{v_p}\cdot\boldsymbol{v_p}>}
\, .
\label{eq:dc:7}
\end{equation} 
The brackets in (\ref{eq:dc:7}) denote an average over time and radii. In
contrast to the definition introduced by \cite{christensen06}, our
modified Rossby number relies on a convective length-scale and
not on the mean half wavelength of the total flow (see also 
the Appendix).

The magnetic field strength is measured by the dimensionless Lorentz number,
\(Lo=B_{\mathrm{rms}}/(\sqrt{\varrho\mu}\Omega L)\), and the classical Elsasser
number \(\Lambda=B_{\mathrm{rms}}^2/{\Omega\varrho\mu\eta}\). They are related 
through \(\Lambda=Lo^2\,Pm/E\). Moreover, following \cite{christensen06}, we 
characterize the geometry of the magnetic field by the relative dipole field 
strength, \(f_\mathrm{dip}\), which is defined as the ratio of the average 
field strength of the dipole field to the field strength in harmonic 
degrees \(\ell=1,\cdots, 12\) at the outer boundary.

A non-dimensional measure for the heat transport is given by the Nusselt 
number, \(Nu\), defined as the ratio of the total heat flow to the 
conducted heat flow, 
\(Q_\mathrm{cond}=4\pi r_o r_i\varrho c \kappa\Delta T/L\) with the heat 
capacity \(c\).
 
\section{Mean-Field Analysis}
We present dynamo models obtained by direct numerical simulations (DNS) and do 
not use the mean-field formalism \citep{steenbeck66,moffatt,raedler80} to 
predict any dynamo action. However, the mean-field approach provides useful
theoretical concepts and mathematical tools to interpret numerical dynamo 
models. It is usually set up by splitting the velocity and the magnetic field 
in mean and fluctuating parts, \(\vec{v}=\overline{\vec{V}}+\vec{v}'\) and
\(\vec{B}=\overline{\vec{B}}+\vec{b}'\). Mean quantities are here denoted by 
an overbar and defined as azimuthal averages. The objective of mean-field 
theory is to predict the evolution of the averaged or mean 
magnetic field, \(\overline{\vec{B}}\), which is in general determined by the 
dynamo equation, 
\begin{equation}
  \frac{\partial{\overline{\vec{B}}}}{\partial t}
  =\nabla\times(\vec{\mathcal{E}}+\overline{\vec{V}}\times\overline{\vec{B}}
  -\eta\nabla\times\overline{\vec{B}}).
  \label{eq:mf:2}
\end{equation}  
The crucial new term in (\ref{eq:mf:2}) compared to the induction equation is 
the mean electromotive force, 
\(\vec{\mathcal{E}}=\overline{\vec{v}'\times\vec{b}'}\). It is a  
functional of \(\vec{v},\overline{\vec{V}}\), and \(\overline{\vec{B}}\), which
is affin-linear in \(\overline{\vec{B}}\).
Provided that there is no small-scale dynamo action, \(\vec{\mathcal{E}}\) is 
also homogeneous in \(\overline{\vec{B}}\). If, moreover, the mean 
electromotive force depends only instantaneously and locally 
on \(\overline{\vec{B}}\), it may be parameterized in terms of the mean field 
and its first derivatives,
\begin{equation}
  \vec{\mathcal{E}}=-\tens{\alpha}\bm-\vec{\gamma}\times\bm-\tens{\beta}(\nabla\times\bm)-\tens{\delta}\times(\nabla\times\bm)-\tens{\kappa}\nabla\bm.
\label{eq:mf:4}
\end{equation} 
The parameters, also known as mean-field coefficients, are vectors 
(\(\vec{\gamma}\) and \(\vec{\delta}\)), symmetric tensors of second 
rank (\(\tens{\alpha}\) and \(\tens{\beta}\)) and a third rank tensor 
(\(\tens{\kappa}\)). They depend only on the velocity field and the magnetic 
diffusivity, but not (explicitly) on the magnetic field.  A physical 
interpretation of the mean-field coefficients is given by 
\cite{raedler95}. The \(\alpha\)-tensor
describes the classical \(\alpha-\)effect \citep[see also][]{parker55}, 
\(\vec{\gamma}\) is associated with an advective transport of the mean 
magnetic field, \(\beta\) can be interpreted as a turbulent 
diffusivity, \(\delta\) may contribute to a shear-current effect, an inductive
effect first noted by \cite{raedler69a,raedler69b}, whereas 
the \(\kappa\)-terms are more difficult to interpret. The altogether 27 
independent mean-field coefficients were determined for several numerical 
models with the help of the so-called test-field method 
\citep{schrinner05,schrinner07}. 

A comparison between DNS and mean-field calculations revealed that 
parameterization (\ref{eq:mf:4}) is indeed reliable for a wide class of 
dynamo models \citep{schrinner11b,schrinner11c}. An important limitation is 
the intrinsic kinematic character of the mean-field approach followed here. 
In general, it is therefore only applicable to dynamo models which can be 
reproduced kinematically and belong to the so-called kinematically stable 
regime identified by \cite{schrinner10a}. 

For a mean-field analysis of some of our dynamo models derived from 
(\ref{eq:dc:2})-(\ref{eq:dc:6}), we solve the dynamo equation as an eigenvalue 
problem 
\begin{equation}
  \sigma\bm=\nabla\times\tens{D}\bm\ ,
  \label{eq:mf:6}
\end{equation}    
with the time-averaged dynamo operator \(\tens{D}\) defined through 
\begin{equation}
\tens{D}\bm=\vm\times\bm-\tens{\alpha}\bm-\vec{\gamma}\times\bm-\tens{\beta}(\nabla\times\bm)-\tens{\delta}\times(\nabla\times\bm)-\tens{\kappa}\nabla\bm.
\label{eq:mf:8}
\end{equation}
On average, the evolution of the mean field is then proportional to 
\(\vec{b}^i\exp{(\sigma_i t)}\) with eigenvectors \(\vec{b}^i\) and 
eigenvalues \(\sigma_i\). More details about the eigenvalue calculation
are given in \cite{schrinner10b}.

\section{Results}
\label{sec:res}
We present results from 72 dynamo models obtained by varying all four 
dimensionless control parameters, the aspect ratio of the shell as well as 
the mechanical boundary conditions. Table \ref{tab:2} is divided in three 
sections for three different types of mechanical boundary conditions and 
provides the control parameters and the output parameters introduced above for 
each model. Within each section, the models are listed in the order of 
increasing local Rossby number. For some models with stress-free mechanical 
boundary conditions, two different solutions coexist depending on the initial 
conditions for the magnetic field. Thus, these models appear in pairs and are
labeled by the letter `d' (dipolar) if the initial magnetic field was strong 
and by `m' (multipolar) if the simulations were started from a weak magnetic 
field.     

\begin{figure}
  \centering
  \includegraphics{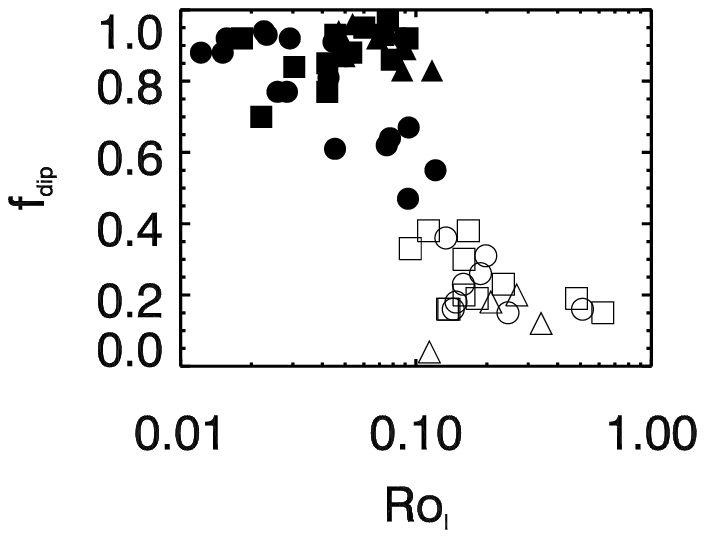}
  \includegraphics{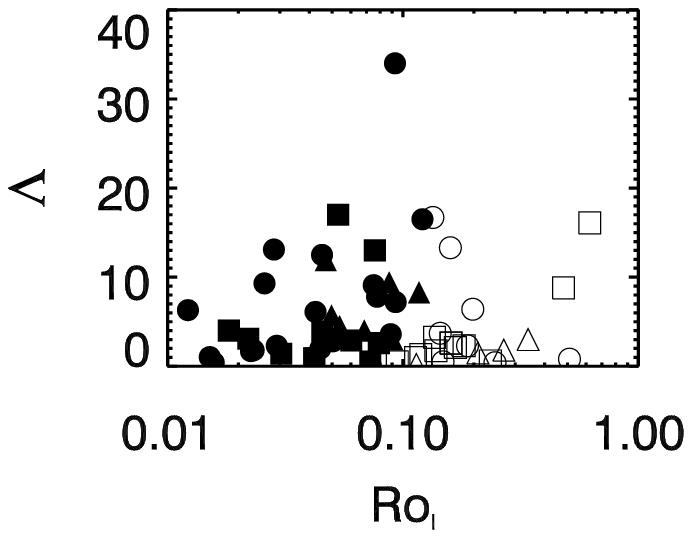}
  \caption{Left: Relative dipole field strength versus the local Rossby number 
  for all models from Table \ref{tbl-1} apart from those with stress-free 
  boundary conditions belonging to the multipolar branch (m-models). 
  Filled symbols stand for models dominated by a dipole field, open symbols 
  denote multipolar models. The symbol shape indicates different types of 
  mechanical boundary conditions: circles mean no-slip conditions at both 
  boundaries, triangles are models with a rigid inner and and a stress-free 
  outer boundary, and squares stand for models with stress-free
  conditions at both boundaries. Right: Elsasser number versus local Rossby 
  number for the same sample of models. 
} 
  \label{fig1}
\end{figure}

\begin{figure}
  \centering
  \includegraphics{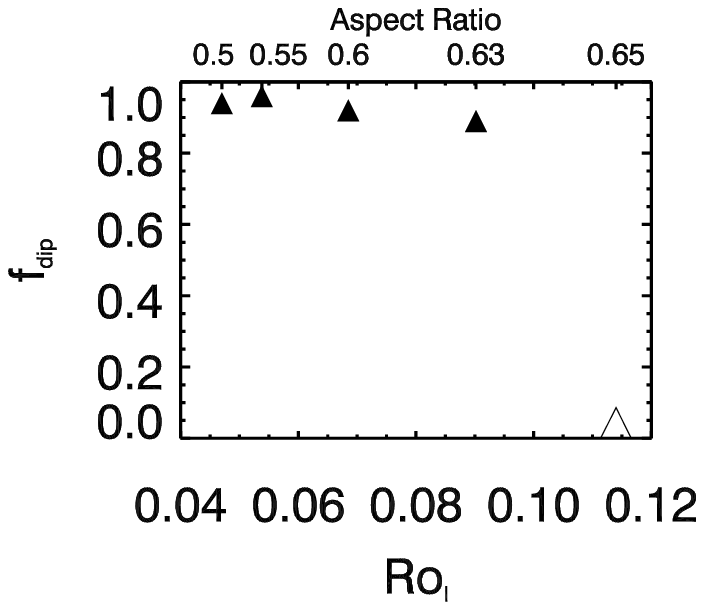}
  \includegraphics{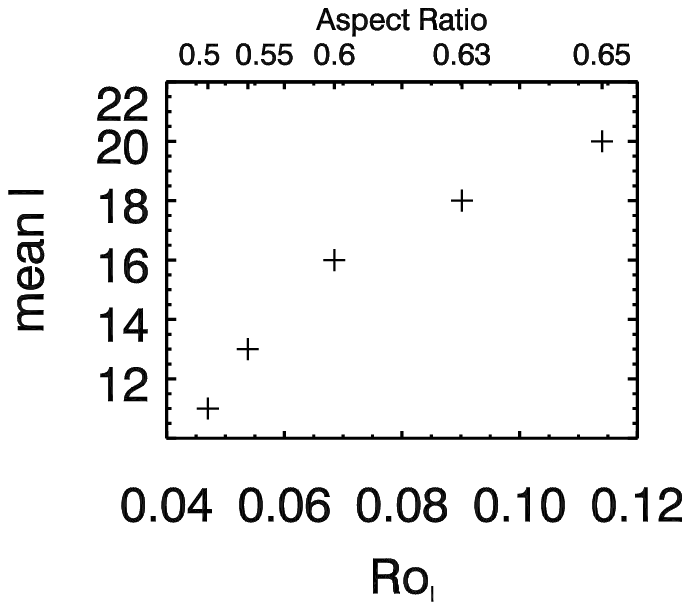}
  \caption{The relative dipole field strength for models 29, 31, 32, 34 (filled
           triangles) and model 35 (open triangle) and the mean 
           harmonic degree \(\overline{\ell}_p\) of the poloidal velocity 
           field (right panel) versus the local Rossby number. The aspect 
           ratio (upper x-axis) has been gradually increased for this sequence 
           of models leading to larger \(\overline{\ell}_p\) and thus also to 
           larger \(Ro_\ell\). Model 35 has undergone a transition from the 
           dipolar to the multipolar dynamo regime and the dipole field 
           strength has dropped drastically. 
  } 
  \label{fig1a}
\end{figure}

\cite{kutzner02} identified a dipolar and a multipolar dynamo regime.
Both regimes can be recovered here if the models are ordered by the local 
Rossby number introduced in section \ref{out_param} 
\citep[see also][]{christensen06}. Figure~\ref{fig1} (left panel)
shows two distinct dynamo regimes: models with a dipole dominated magnetic 
field at low Rossby numbers (filled symbols) and models with a more 
multipolar magnetic field at 
higher Rossby numbers (open symbols). A fairly sharp transition between both 
regimes occurs at \(Ro_\ell\approx 0.1\). The exact value may depend weakly on 
the choice of the mechanical boundary conditions and seems to be closer 
to \(Ro_\ell\approx0.12\) for models with rigid boundaries. The right panel 
of Fig. \ref{fig1} illustrates that models of both regimes differ in their 
field topology, but not necessarily in their field strength.  We emphasize 
that only the definition  of a local Rossby number based on a convective 
length scale (i.e. not taking into account the zonal flow) enables us to 
extend the well known dipole-multipole partition to models with different 
aspect ratios and boundary conditions. This is most clearly demonstrated by 
the sequence of models 29, 31, 32, 34 and 35 from Table \ref{tab:2} which is 
depicted in Fig. \ref{fig1a}. For these models, the Ekman number, the 
Rayleigh number (normalized by its critical value), and the Prandtl numbers 
were kept constant whereas the aspect ratio was progressively increased. 
The thinner the convection zone, the smaller the convective length scales, 
and the mean harmonic degree of the poloidal velocity field increases 
systematically from \(\overline{\ell}_p=11\) for model 29 with \(\chi=0.5\) to 
\(\overline{\ell}_p=20\) for model 35 with \(\chi=0.65\). Consequently, 
\(Ro_\ell\) grows from \(Ro_\ell=4.70\times 10^{-2}\) for model 29 to 
\(Ro_\ell=1.14\times 10^{-1}\) for model 35, which already belongs to the 
multipolar regime. The decrease of the convective length scales caused a 
transition towards the multipolar regime and is adequately measured by the 
local Rossby number introduced here. If, however, \(\overline{\ell}_p\) were 
derived from the \textit{total} velocity field, neither \(\overline{\ell}_p\) 
nor \(Ro_\ell\) would increase in the same way, owing to a major contribution 
of the mean zonal flow to the kinetic energy density in model 35. The Rossby
number criterion would then fail to predict the observed 
transition between both dynamo regimes. Note that decreasing the Ekman
number from \(E=10^{-3}\) to \(E=3\, 10^{-4}\) in model 30 leads again to a 
dipolar field despite the high aspect ratio of \(\chi=0.65\) and that we find 
a multipolar magnetic field for model 37 at lower aspect ratio, \(\chi=0.6\), 
but higher Rayleigh number. Both examples illustrate that the transition is 
indeed characterized by \(Ro_\ell\). However, the definition of \(Ro_\ell\) 
is empirically motivated, there may be other ways to define a relevant Rossby 
number to distinguish both dynamo regimes; some of them are discussed in 
Appendix A. 

Moreover, Fig. \ref{fig3} demonstrates that the relative dipole field 
strength does not simply depend on the magnetic Reynolds number. 
The distance from the dynamo threshold of a model does not determine its 
field topology.


An exception to the Rossby number criterion for the dipolarity of
the magnetic field in dynamo models occurs in the case of stress 
free boundary conditions. Simulations started from a weak magnetic 
field (m-models) are not simply dipole dominated, independent of their local 
Rossby number. These models left out in Fig. \ref{fig1} were included in 
Fig. \ref{fig2} and are represented by diamonds. As apparent from 
Fig. \ref{fig2}, dynamo models with stress free mechanical boundary conditions 
form two branches, an upper branch of dipolar models for \(Ro_\ell< 0.1\) and 
a lower branch of models with a less constrained field geometry 
\citep[see also][]{busse09,simitev09}. We refer to them as the `dipolar' 
and the `multipolar' branch instead of the `weak-field' and the `strong-field' 
branch to avoid confusion with corresponding notions introduced in the limit 
of vanishing viscosity and inertia in the context of planetary dynamos. 
Nevertheless, the saturated magnetic field strength, as 
measured by the Lorentz number, is always larger for dipolar than for 
multipolar models. For \(Ro_\ell>0.1\), both branches coincide 
and their distinction is no longer meaningful (see Fig.\ref{Ra_Els_branch}). 
The region of bistability in parameter space, however, does not solely depend 
on the local Rossby number, but also on the magnetic Prandtl number 
\citep{busse09}; for models 40, 42, and 44, a weak-field branch does not 
exist.     
        
There is a strong correlation between the topology and the time dependence of 
the magnetic field in dynamo models. Sudden polarity reversals or oscillations 
of the magnetic field do not occur in dipole dominated models in the low 
Rossby number regime. Conversely, reversals and oscillations are frequent in 
non-dipolar models with \(Ro_\ell>0.1\) as well as in models with lower
local Rossby numbers with stress-free boundary conditions belonging to the 
multipolar branch. Whether non-dipolar models exhibit fairly coherent 
oscillations or irregular reversals of the magnetic field strongly depends on 
the magnetic Reynolds number. Coherent oscillatory solutions 
of the induction equation are most clearly visible in so-called butterfly 
diagrams; contours of the azimuthally averaged radial magnetic field at the 
outer boundary are plotted versus time and colatitude. Figure \ref{fig5} gives 
two examples. The left panel shows a very coherent dynamo wave  at
\(Rm=102\) (model 45m), whereas the butterfly diagram  at \(Rm=258\) on 
the right-hand side (model 38) is much less periodic and a cycle period cannot 
be identified. Dynamo models (in the non-dipolar regime) at higher magnetic 
Reynolds number exhibit even less temporal coherence. Following this somewhat 
arbitrary and qualitative criterion, we find that non-dipolar dynamos of our 
sample with \(Rm\lesssim 200\) generate magnetic fields which vary 
periodically in time. The lower the magnetic Reynolds number, the more 
coherent is the time variability of the magnetic field.  
\begin{figure}
  \centering
  \includegraphics{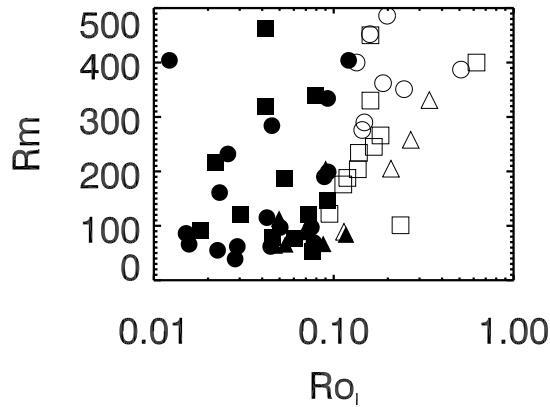}
  \caption{The magnetic Reynolds number versus the local Rossby number 
  for all models from Table \ref{tbl-1}. For the meaning of 
  the symbol style we refer to the caption of Fig. \ref{fig1}. 
}
  \label{fig3}
\end{figure} 
\begin{figure}[h]
  \centering
  \includegraphics{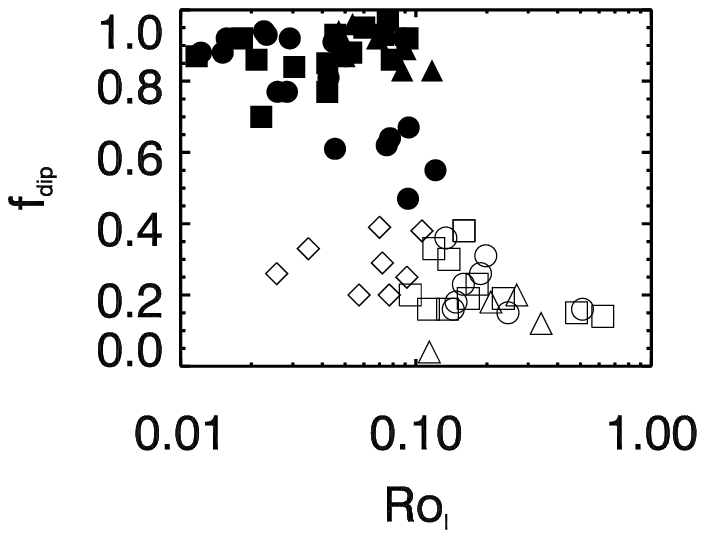}
  \caption{The relative dipole field strength versus the local Rossby number
  as in Fig. \ref{fig1} but here with the m-models included (diamonds). All 
  other notations as in Fig. \ref{fig1}.}
  \label{fig2}
\end{figure} 
A third property is intimately linked to the topology and the time dependence
of the magnetic field. \cite{schrinner10a} looked at
the evolution of a passive vector field described by a second induction 
equation which was solved simultaneously with equations 
(\ref{eq:dc:2})-(\ref{eq:dc:6}). The magnetic `tracer-field' was advanced at
each time-step with the self-consistently determined velocity field but did 
not contribute to the Lorentz force. A similar experiment had been performed 
by \cite{cattaneo09b} and \cite{tilgner08}, too. \cite{schrinner10a} found 
that the tracer field grows exponentially for multipolar and reversing models, 
whereas it remains kinematically stable for dipolar dynamos in the low Rossby 
number regime. This relation can be extended to the sample of models studied 
here and serves to distinguish both dynamo classes. Dipolar models represented 
by filled symbols in Fig. \ref{fig2} are kinematically stable, all multipolar 
models represented by open symbols are kinematically unstable.
\clearpage  
\begin{deluxetable}{ccccccccccc}
\tabletypesize{\scriptsize}
\tablecaption{Overview of the simulations carried out, ordered with respect to 
their modified Rossby number and their mechanical boundary conditions.
\label{tbl-1}}
\tablewidth{0pt}
\tablehead{
\colhead{Model} & \colhead{\(E\)} & \colhead{\(Ra\)} & \colhead{\(Pm\)} & 
\colhead{\(Pr\)} & \colhead{\(\chi\)} & \colhead{\(\overline{\ell}_p\)} & \colhead{\(Ro_\ell\)} & 
\colhead{\(f_\mathrm{dip}\)} & \colhead{\(Rm\)} & \colhead{\(Lo\)} 
}
\startdata
\multicolumn{11}{c}{Rigid Boundary Conditions}\\
 1   & \(3\times 10^{-5}\) & 510& 3 & 1 & 0.35 & 14 & \(7.37\times 10^{-3}\)& 0.88 & 161 & \(4.92\times 10^{-3}\)\\
 2   & \(3\times 10^{-5}\) & 510& 4 & 1 & 0.35 & 14 & \(7.88\times 10^{-3}\)& 0.91 & 232 & \(8.02\times 10^{-3}\)\\ 
 3   & \(1\times 10^{-3}\) & 100& 5 & 1 & 0.35 &  5  & \(1.22\times 10^{-2}\)& 0.88  & 39&\(3.54\times 10^{-2}\)\\
 4   & \(1\times 10^{-4}\) & 334& 2 & 1 & 0.35 & 11 & \(1.51\times 10^{-2}\)& 0.88 &  86 & \(7.18\times 10^{-3}\)\\
 5   & \(3\times 10^{-4}\) & 195& 3 & 1 & 0.35 &  8 & \(1.57\times 10^{-2}\)& 0.92 &  66 & \(6.53\times 10^{-3}\)\\
 6   & \(3\times 10^{-4}\) & 243& 2 & 1 & 0.35 &  9 & \(2.26\times 10^{-2}\)& 0.94 &  55 & \(1.61\times 10^{-3}\)\\
 7   & \(3\times 10^{-5}\) & 750& 1 & 1 & 0.35 & 25 & \(2.32\times 10^{-2}\)& 0.93 &  97 & \(7.35\times 10^{-2}\)\\ 
 8   & \(1\times 10^{-3}\) & 150& 5 & 1 & 0.40 &  6 & \(2.58\times 10^{-2}\)& 0.77 &  69 & \(4.32\times 10^{-3}\)\\
 9   & \(1\times 10^{-3}\) & 136& 5 & 1 & 0.45 &  7 & \(2.83\times 10^{-2}\)& 0.77 &  62 & \(5.11\times 10^{-2}\)\\
 10  & \(3\times 10^{-4}\) & 285& 2 & 1 & 0.35 & 10 & \(2.91\times 10^{-2}\)& 0.92 &  62 & \(1.84\times 10^{-2}\)\\
 11  & \(3\times 10^{-4}\) & 375& 3 & 1 & 0.35 & 12 & \(4.25\times 10^{-2}\)& 0.81 & 115 & \(2.47\times 10^{-2}\)\\
 12  & \(3\times 10^{-4}\) & 375& 1.5 & 1 & 0.35 & 11 & \(4.46\times 10^{-2}\)& 0.91 &  62 & \(1.98\times 10^{-2}\)\\
 13  & \(3\times 10^{-4}\) & 330& 9 & 3 & 0.35 & 15 & \(4.53\times 10^{-2}\)& 0.61 &  284 & \(2.04\times 10^{-2}\)\\
 14  & \(3\times 10^{-4}\) & 330& 3 & 3 & 0.35 & 16 & \(5.04\times 10^{-2}\)& 0.87 &  97 & \(1.69\times 10^{-2}\)\\
 15  & \(1\times 10^{-3}\) & 100& 7 & 1 & 0.65 & 16 & \(7.51\times 10^{-2}\)& 0.62 & 107 & \(3.60\times 10^{-2}\)\\
 16  & \(1\times 10^{-3}\) & 100& 6 & 1 & 0.65 & 16 & \(7.76\times 10^{-2}\)& 0.64 &  93 & \(3.61\times 10^{-2}\)\\
 17  & \(1\times 10^{-4}\) & 1117& 1.5 & 0.67 & 0.35 & 22 & \(8.87\times 10^{-2}\)& 0.92 & 190 & \(1.54\times 10^{-2}\)\\
 18  & \(1\times 10^{-3}\) & 400& 10 & 1 & 0.35 & 9 & \(9.25\times 10^{-2}\)& 0.47 &  334 & \(5.81\times 10^{-2}\)\\
 19  & \(3\times 10^{-4}\) & 630& 3 & 1 & 0.35 & 15 & \(9.31\times 10^{-2}\)& 0.67 & 199 & \(2.68\times 10^{-2}\)\\
 20  & \(3\times 10^{-4}\) & 810& 5 & 1 & 0.35 & 16 & \(1.21\times 10^{-1}\)& 0.55 & 404 & \(3.15\times 10^{-2}\)\\
 21  & \(1\times 10^{-3}\) & 450& 10& 1 & 0.35 & 11 & \(1.34\times 10^{-1}\)& 0.36 & 401 & \(4.08\times 10^{-2}\)\\
 22  & \(3\times 10^{-4}\) & 810& 3 & 1 & 0.35 & 16 & \(1.44\times 10^{-1}\)& 0.16 & 277 & \(1.93\times 10^{-2}\)\\
 23  & \(3\times 10^{-4}\) & 750& 3 & 1 & 0.35 & 16 & \(1.48\times 10^{-1}\)& 0.18 & 290 & \(7.14\times 10^{-2}\)\\
 24  & \(1\times 10^{-3}\) & 500& 10& 1 & 0.35 & 11 & \(1.59\times 10^{-1}\)& 0.23 & 453 & \(3.65\times 10^{-3}\)\\
 25  & \(3\times 10^{-4}\) & 1050& 3& 1 & 0.35 & 16 & \(1.88\times 10^{-1}\)& 0.26 & 363 & \(1.52\times 10^{-2}\)\\
 26  & \(3\times 10^{-4}\) & 1250& 3& 0.3 & 0.35 & 13 & \(1.98\times 10^{-1}\)& 0.31 & 486 & \(2.52\times 10^{-2}\)\\
 27  & \(1\times 10^{-4}\) & 1117& 1.5 & 0.67 & 0.5 & 33 & \(2.46\times 10^{-1}\)& 0.15 & 351 & \(5.06\times 10^{-3}\)\\
 28  & \(3\times 10^{-4}\) & 2970& 1& 0.3 & 0.35 & 14 & \(5.10\times 10^{-1}\)& 0.16 & 387 & \(1.55\times 10^{-2}\)\\
\tableline
\multicolumn{11}{c}{Mixed Boundary Conditions}\\
29   & \(1\times 10^{-3}\) & 125& 5& 1 & 0.50 & 11 & \(4.70\times 10^{-2}\)& 0.94 &  67 & \(4.88\times 10^{-2}\)\\
30   & \(3\times 10^{-4}\) & 120& 5& 1 & 0.65 & 26 & \(4.97\times 10^{-2}\)& 0.87 & 100 & \(1.86\times 10^{-2}\)\\
31   & \(1\times 10^{-3}\) & 110& 5& 1 & 0.55 & 13 & \(5.38\times 10^{-2}\)& 0.96 &  64 & \(2.99\times 10^{-2}\)\\
32   & \(1\times 10^{-3}\) & 105& 5& 1 & 0.60 & 16 & \(6.85\times 10^{-2}\)& 0.92 &  66 & \(2.84\times 10^{-2}\)\\
33   & \(1\times 10^{-3}\) & 125& 5& 1 & 0.60 & 16 & \(8.74\times 10^{-2}\)& 0.83 &  84 & \(4.34\times 10^{-2}\)\\
34   & \(1\times 10^{-3}\) & 105& 5& 1 & 0.63 & 18 & \(9.02\times 10^{-2}\)& 0.89 &  78 & \(2.47\times 10^{-2}\)\\
35   & \(1\times 10^{-3}\) & 100& 5& 1 & 0.65 & 20 & \(1.14\times 10^{-1}\)& 0.04 &  89 & \(6.63\times 10^{-3}\)\\
36   & \(1\times 10^{-3}\) & 150& 5& 1 & 0.60 & 16 & \(1.17\times 10^{-1}\)& 0.83 & 113 & \(4.09\times 10^{-2}\)\\
37   & \(1\times 10^{-3}\) & 175& 5& 1 & 0.60 & 16 & \(2.08\times 10^{-1}\)& 0.18 & 205 & \(1.71\times 10^{-2}\)\\ 
38   & \(1\times 10^{-3}\) & 200& 5& 1 & 0.60 & 16 & \(2.68\times 10^{-1}\)& 0.20 & 258 & \(1.87\times 10^{-2}\)\\
39   & \(1\times 10^{-3}\) & 250& 5& 1 & 0.60 & 16 & \(3.40\times 10^{-1}\)& 0.12 & 331 & \(2.55\times 10^{-2}\)\\
\tableline
\multicolumn{11}{c}{Stress-Free Boundary Conditions}\\
 40  & \(1\times 10^{-4}\) & 365& 2& 1 & 0.35 & 12 & \(1.82\times 10^{-2}\)& 0.92 & 92 & \(5.69\times 10^{-3}\)\\
 41d & \(3\times 10^{-5}\) & 600& 1& 1 & 0.35 & 16 & \(2.11\times 10^{-2}\)& 0.86 & 140 & \(3.82\times 10^{-3}\)\\
 41m & \(3\times 10^{-5}\) & 600& 1& 1 & 0.35 & 16 & \(2.56\times 10^{-2}\)& 0.26 & 144 & \(2.35\times 10^{-3}\)\\
 42  & \(1\times 10^{-4}\) & 375& 4& 1 & 0.35 & 13 & \(2.20\times 10^{-2}\)& 0.70 & 216 & \(8.80\times 10^{-3}\)\\
 43d & \(1\times 10^{-4}\) & 462& 2& 1 & 0.35 & 16 & \(3.04\times 10^{-2}\)& 0.84 & 121 & \(8.44\times 10^{-3}\)\\
 43m & \(1\times 10^{-4}\) & 462& 2& 1 & 0.35 & 16 & \(3.49\times 10^{-2}\)& 0.33 & 146 & \(4.91\times 10^{-3}\)\\
 44  & \(1\times 10^{-4}\) & 750& 6& 1 & 0.35 & 17 & \(4.20\times 10^{-2}\)& 0.77 & 462 & \(2.50\times 10^{-2}\)\\
 45d & \(1\times 10^{-4}\) & 586& 1& 1 & 0.35 & 18 & \(4.54\times 10^{-2}\)& 0.93 &  78 & \(9.71\times 10^{-3}\)\\
 45m & \(1\times 10^{-4}\) & 586& 1& 1 & 0.35 & 18 & \(5.73\times 10^{-2}\)& 0.20 &  102 & \(5.60\times 10^{-3}\)\\
 46d & \(1\times 10^{-4}\) & 749& 2& 1 & 0.35 & 18 & \(5.30\times 10^{-2}\)& 0.88 &  187 & \(1.40\times 10^{-2}\)\\
 46m & \(1\times 10^{-4}\) & 749& 2& 1 & 0.35 & 18 & \(7.00\times 10^{-2}\)& 0.39 &  244 & \(8.12\times 10^{-3}\)\\
 47d & \(1\times 10^{-4}\) & 750& 4& 1 & 0.35 & 17 & \(4.20\times 10^{-2}\)& 0.85 & 320 & \(2.06\times 10^{-2}\)\\
 47m & \(1\times 10^{-4}\) & 750& 4& 1 & 0.35 & 20 & \(7.20\times 10^{-2}\)& 0.29 & 460 & \(9.89\times 10^{-3}\)\\
 48d & \(1\times 10^{-4}\) & 750& 0.75& 1 & 0.35 & 19 & \(6.03\times 10^{-2}\)& 0.95& 77& \(1.11\times 10^{-2}\)\\
 48m & \(1\times 10^{-4}\) & 750& 0.75& 1 & 0.35 & 18 & \(7.70\times 10^{-2}\)& 0.20 & 100 & \(7.02\times 10^{-3}\)\\
 49d & \(3\times 10^{-4}\) & 510& 2& 1 & 0.35 & 13 & \(7.26\times 10^{-2}\)& 0.93 & 120 & \(2.08\times 10^{-2}\)\\
 49m & \(3\times 10^{-4}\) & 510& 2& 1 & 0.35 & 12 & \(9.15\times 10^{-2}\)& 0.25 & 155 & \(1.09\times 10^{-2}\)\\
 50d  & \(1\times 10^{-4}\) & 750& 0.5& 1 & 0.35 & 12 & \(7.60\times 10^{-2}\)& 0.97 & 53 & \(1.03\times 10^{-2}\)\\
 50m  & \(1\times 10^{-4}\) & 750& 0.5& 1 & 0.35 & 12 & \(8.20\times 10^{-2}\)& 0.21 & 70 & \(6.32\times 10^{-3}\)\\
 51d & \(1\times 10^{-4}\) & 750& 2& 0.3 & 0.35 & 15 & \(7.90\times 10^{-2}\)& 0.86 & 340 & \(2.55\times 10^{-2}\)\\
 51m & \(1\times 10^{-4}\) & 750& 2& 0.3 & 0.35 & 15 & \(1.06\times 10^{-1}\)& 0.38 & 450 & \(1.52\times 10^{-2}\)\\
 52d & \(1\times 10^{-4}\) & 1110& 1& 1 & 0.35 & 20 & \(9.20\times 10^{-2}\)& 0.92 & 146 & \(1.61\times 10^{-2}\)\\
 52m & \(1\times 10^{-4}\) & 1110& 1& 1 & 0.35 & 20 & \(1.19\times 10^{-1}\)& 0.33 & 188 & \(1.13\times 10^{-2}\)\\
 53  & \(3\times 10^{-4}\) & 510& 1.5& 1 & 0.35 & 12 & \(9.46\times 10^{-2}\)& 0.20 & 122 & \(1.01\times 10^{-2}\)\\
 54  & \(1\times 10^{-4}\) & 1333& 2& 0.3 & 0.35 & 15 & \(1.60\times 10^{-1}\)& 0.38 & 450 & \(1.52\times 10^{-2}\)\\
 55  & \(1\times 10^{-4}\) & 1000& 1& 1 & 0.35 & 20 & \(1.13\times 10^{-1}\)& 0.16 & 176 & \(9.83\times 10^{-3}\)\\
 56  & \(1\times 10^{-4}\) & 1200& 1& 1 & 0.35 & 21 & \(1.36\times 10^{-1}\)& 0.16 & 204 & \(1.18\times 10^{-2}\)\\
 57  & \(1\times 10^{-4}\) & 1372& 1& 1 & 0.35 & 19 & \(1.38\times 10^{-1}\)& 0.30 & 234 & \(1.32\times 10^{-2}\)\\
 58  & \(1\times 10^{-4}\) & 1757& 1& 0.3 & 0.35 & 15 & \(1.60\times 10^{-1}\)& 0.38 & 330 & \(1.61\times 10^{-2}\)\\
 59  & \(1\times 10^{-4}\) & 1497& 1& 1 & 0.35 & 21 & \(1.67\times 10^{-1}\)& 0.19 & 245 & \(1.45\times 10^{-2}\)\\
 60  & \(1\times 10^{-4}\) & 1627& 1& 1 & 0.35 & 22 & \(1.82\times 10^{-1}\)& 0.23 & 266 & \(1.52\times 10^{-2}\)\\
 61  & \(1\times 10^{-4}\) & 1823& 0.3& 1 & 0.35 & 22 & \(2.35\times 10^{-1}\)& 0.19 & 101 & \(1.45\times 10^{-2}\)\\
 62  & \(3\times 10^{-4}\) & 2000&  2& 0.3 & 0.35 & 15 & \(4.81\times 10^{-1}\)& 0.15 & 658 & \(3.63\times 10^{-2}\)\\
\enddata
\tablecomments{Some of the models in Table \ref{tbl-1} were already published 
in \cite{schrinner10a} with a somewhat different definition of 
\(\overline{\ell}_p\) and thus also of \(Ro_\ell\).}
\label{tab:2}
\end{deluxetable}
\clearpage
\begin{figure}
  \centering
  \includegraphics{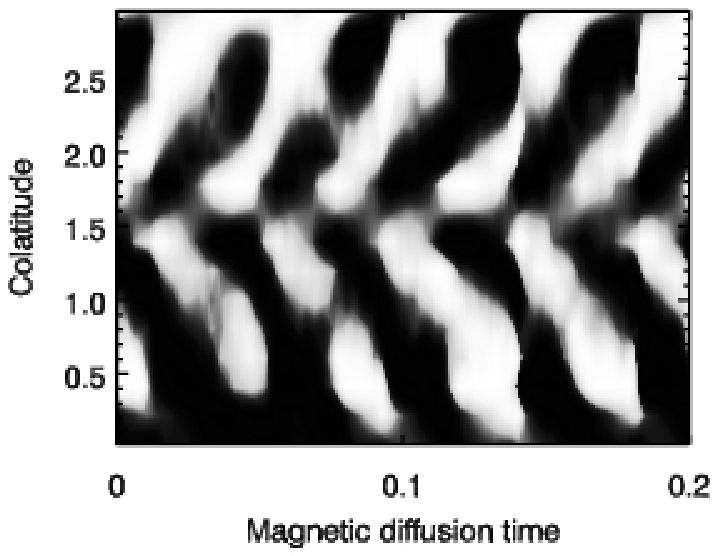}
  \includegraphics{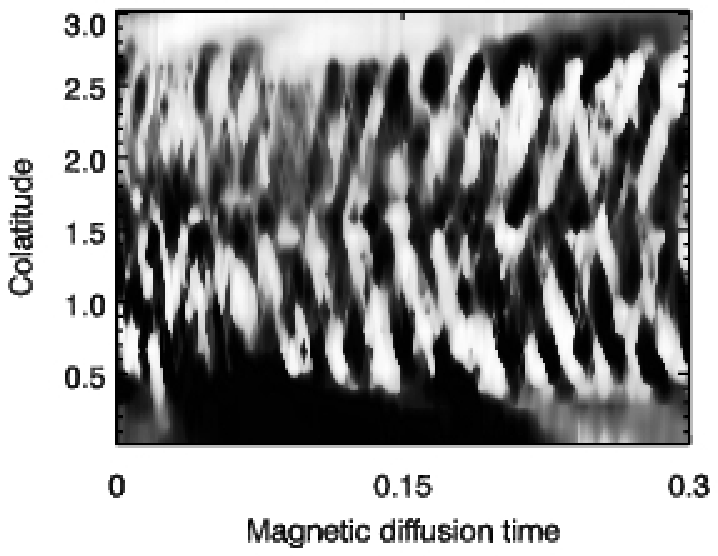}
  \caption{Left panel: Butterfly diagram for model 45m, \(Rm=102\). 
   Right panel: Butterfly diagram for model 38, \(Rm=258\). The contour plots 
   are normalized by their maximum absolute value and the grey-scale coding 
   varies from +1 (black) to -1 (white). }
  \label{fig5}
\end{figure} 
\begin{figure}
  \centering
  \includegraphics{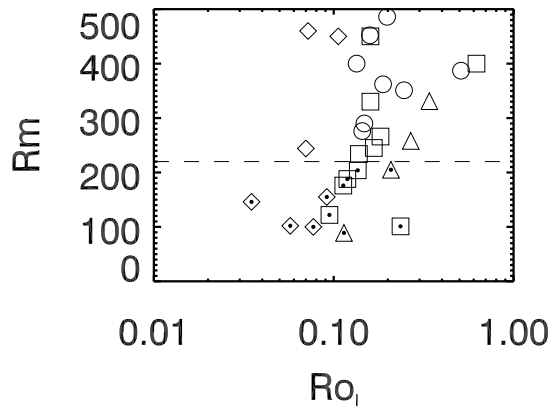}
  \caption{Magnetic Reynolds number versus the local Rossby number for all
   \textit{multipolar} models of Table \ref{tbl-1}. A dot in an open symbol 
   indicates that a coherent dynamo wave was found. The meaning of all other 
   notations is as defined in Fig. \ref{fig1} and Fig. \ref{fig2}.}
  \label{fig4}
\end{figure} 
\clearpage

\section{Discussion}
\subsection{Reason for the dipole breakdown}
What causes the dipolarity of dynamo models far above the dynamo threshold in 
the low Rossby number regime? The Rossby number criterion supported by this 
work predicts a dipole dominated magnetic field, if the typical length scale of
convection, \((\overline{\ell}_p/\pi)\,L\), is at least one order of magnitude 
larger than the Rossby radius, i.e. 
\(0.1\,(\overline{\ell}_p/\pi)\,L>v_\mathrm{rms}/\Omega\). Convection is then 
strongly influenced by rotation and organized in columns parallel to the 
rotation axis. A mean-field analysis reveals that the magnetic field generated 
and maintained by these convective rolls is dominated by only one real dipolar 
eigenmode with approximately zero growth rate. All more structured and 
in general complex overtones (i.e. higher order modes) are highly diffusive. 
The kinematic stability of models at low \(Ro_\ell\) as well as their 
comparatively steady dipole field are a consequence of this single mode 
property \citep{schrinner10a}. However, the reason for the breakdown of the 
dipole field at \(Ro_\ell\approx 0.1\) is at present not well understood. 
We computed the eigenvalue spectra and the eigenmodes of the 
time-averaged dynamo operator \(D\) for a sequence of models with 
increasing \(Ro_\ell\) to gain more insight. The sequence consists of model 29,
models 31--32, and model 34, already introduced in section \ref{sec:res}. 
These are kinematically stable models ($Ro_\ell< 0.1$) with mixed mechanical 
boundary conditions and aspect ratios varying from \(\chi=0.5\) to 
\(\chi=0.63\). Models with the same parameter values and boundary conditions 
but an aspect ratio lower than \(0.5\) do not exhibit any dynamo action. We 
therefore considered in addition model 4 with rigid boundary conditions as an
example of a dynamo model at lower Rossby number.

\begin{figure}
  \centering
  \includegraphics{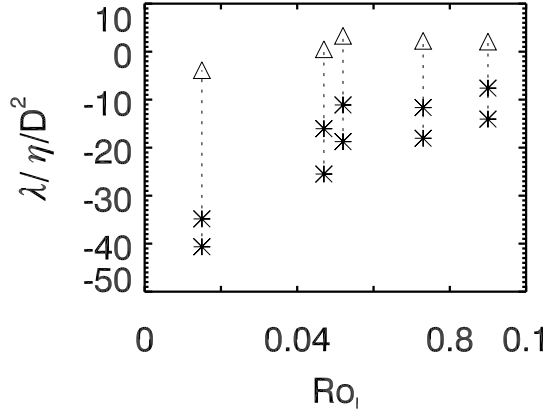}
  \caption{Growth rates of the fundamental mode (triangle) and of the first
  two dipolar overtones (stars) for five models at different Rossby 
  number; in the order of increasing Rossby number, these models are: model 4, 
  model 29, model 31, model 32, and model 34. }
  \label{fig7}
\end{figure}

Figure \ref{fig7} shows the growth rates of the fundamental mode and of 
the first two dipolar overtones versus the local Rossby number for these five 
models. The fundamental modes have on time average approximately zero growth 
rate, as it is expected for saturated dynamos, all overtones are diffusive.  
While there is typically a large gap between the fundamental mode and the 
first overtone for models at low \(Ro_\ell\), both growth rates are much closer
if the Rossby number increases. Quadrupolar modes were omitted in 
Fig. \ref{fig7} because the field realized in the DNS is of purely dipolar 
symmetry for the five examples considered. For completeness, the eigenvalues 
of the first dipolar and quadrupolar modes are listed in Table \ref{tab:4}. 
We emphasize again that the change in the spectra visible in Fig. \ref{fig7} 
is not correlated to an increase of \(Rm\). The magnetic Reynolds number is 
highest for model 4 (\(Rm=86\)) and does not change significantly for the 
following models at larger \(Ro_\ell\).

\begin{deluxetable}{crrrrr}
\tabletypesize{\scriptsize}
\tablecaption{Eigenvalues for the first modes with dipolar (\(\sigma^a\)) and
quadrupolar (\(\sigma^s\)) symmetry for the five models considered in Fig. 
\ref{fig7} in units of \(\eta/L^2\).}
\tablewidth{0pt}
\tablehead{
\colhead{Model} & \colhead{\(\sigma_0^a\)} & \colhead{\(\sigma_1^a\)} & 
 \colhead{\(\sigma_2^a\)} & \colhead{\(\sigma_0^s\)} & 
\colhead{\(\sigma_1^s\)}  
}
\startdata
4&\((-3.87, 0.00)\)&\((-34.8,\pm 10.3)\)&\((-40.6,0.00)\)&\((-6.30,0.00)\)
&\((-28.7,\pm 2.69)\)\\
29&\((0.50,0.00)\)&\((-16.1,0.00)\)&\((-25.5,\pm 4.21)\)&\((-18.3,0.00)\)
&\((-25.5,\pm 6.19)\)\\
31&\((3.30,0.00)\)&\((-11.1,0.00)   \) &\((-18.8,\pm 2.75)\)
&\((-10.3,\pm 1.70)\)&\((-20.0,\pm 4.88)\)\\
32&\((2.25,0.00)\)&\((-11.7,\pm 2.51)\)&\((-18.1,0.00)\) &\((-10.0,\pm 2.50)\)
&\((-17.5,0.00)\)\\
34&\((2.10,0.00)\)&\((-7.60,0.00)\)&\((-14.1,0.00)\)
&\((-9.35,0.00)\)&\((-10.5,0.00)\)\\
\enddata
\label{tab:4}
\end{deluxetable}

\begin{figure}
  \centering
  \includegraphics{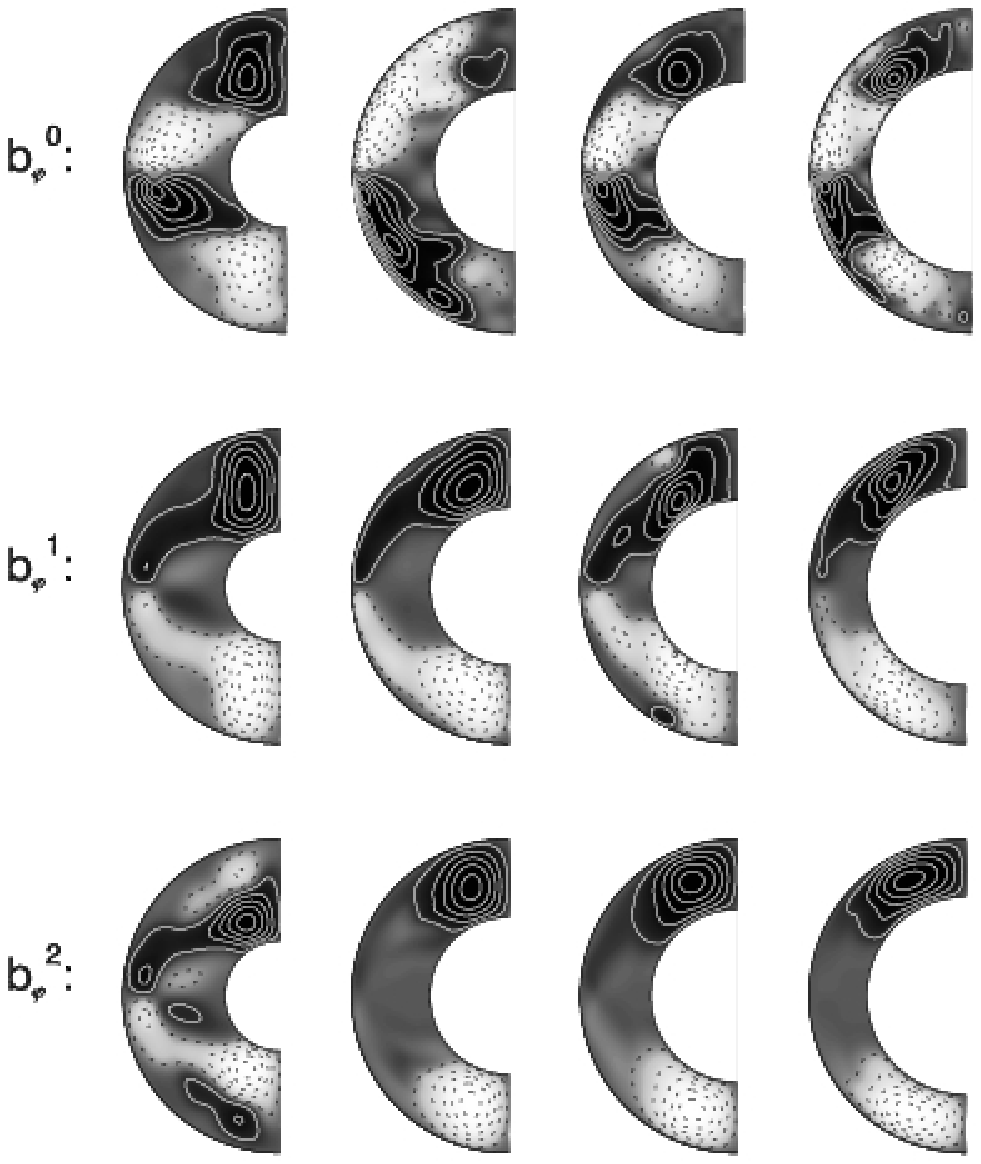}
  \caption{Contour plots of the \(\varphi-\)components of the fundamental 
    mode, \(b_\varphi^0\), and of the first dipolar overtones, 
    \(b_\varphi^1\), \(b_\varphi^2\), for model 4, model 29, model 31, and 
    model 34 (in columns from left to right). The first overtone of model 4 
   (first column) and \(b_\varphi^2\) of model 29 and model 31  are complex 
   and only their real part is shown. The aspect ratio and the local Rossby 
   number increase from model 4 to model 34. Each contour plot is normalized 
   by its maximum absolute value and the gray-scale coding varies 
   from -1 (white) to +1 black. The contour lines correspond to 
   \(\pm 0.1, \pm 0.3, \pm0.5, \pm 0.7\), and \(\pm 0.9\).} 
  \label{fig8}
\end{figure}

At low Rossby numbers, the fundamental mode is well separated from the
following modes (or overtones) which are strongly damped (see Fig.~\ref{fig7}).
We refer to this characteristic as the `single-mode' property. As the 
Rossby number is increased, the average growth rates converge to zero, in 
other words, the eigenvalues of the overtones approach that of the 
leading eigenmode. This appears to result from the fact that the structure of 
the fundamental mode becomes more similar to the following overtones, too. 
Differences between the first eigenmodes are most visible in their 
\(\varphi\)-components. Figure \ref{fig8} shows contour plots of the 
fundamental mode and the first two dipolar eigenmodes for model 4, model 29, 
model 31, and model 34. The axisymmetric azimuthal field of dynamo models at 
low Rossby number is dominated by two flux bundles of inverse polarity close 
to the equatorial plane near the outer shell boundary 
\citep[see also][]{olson99}. These flux patches are visible in the 
\(\varphi\)-components of the fundamental modes, \(b_\varphi^0\), for 
all four models. However, as the aspect ratio and the Rossby number increase, 
the axisymmetric flux concentration at low latitudes becomes less pronounced. 
The fundamental mode for model 34 resembles in this respect its following 
overtones and it is probably this adjustment in the eigenmodes which causes 
the convergence of the eigenvalues for dynamo models at 
\(Ro_\ell\approx 0.1\). 

\begin{figure}
  \centering
  \includegraphics{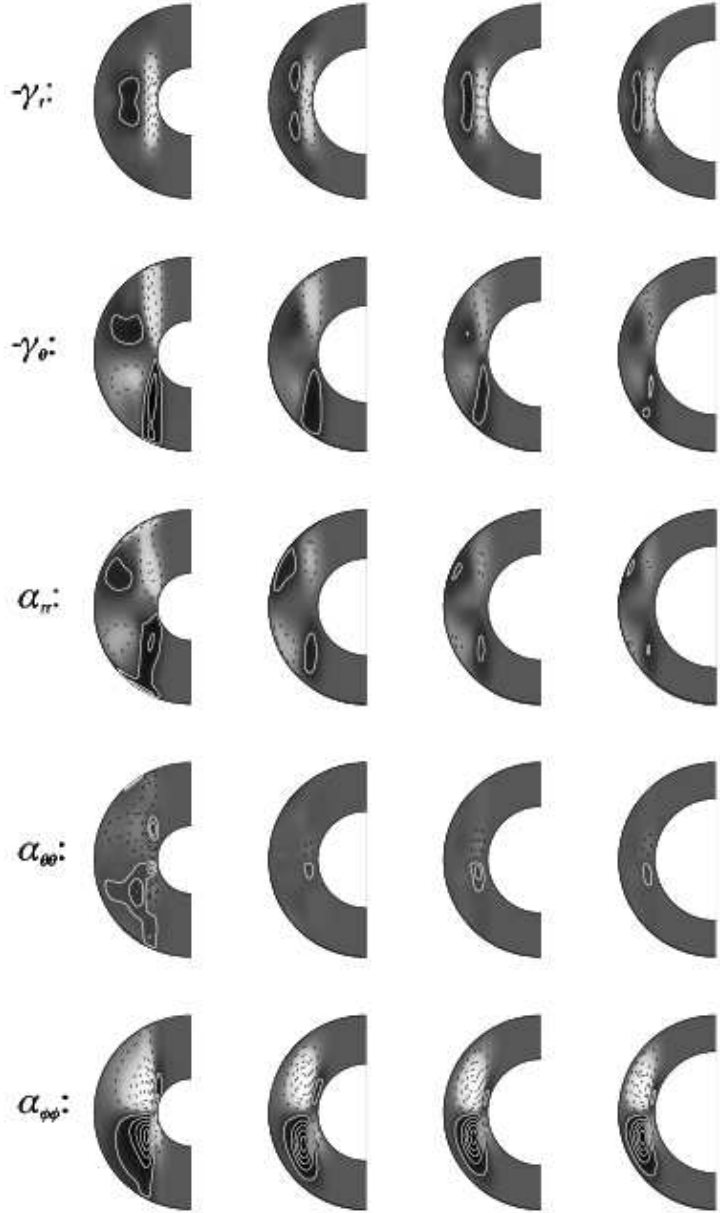}
  \caption{Dynamo coefficients for model 4, model 29, model 31 
   and model 34 (in columns). The mean azimuthal field is 
   advected in radial and latitudinal direction by 
  \(-\gamma_r\) and \(-\gamma_\theta\), and generated from poloidal field by
  \(-\alpha_{rr}\) and \(-\alpha_{\theta\theta}\). For each model, the dynamo
  coefficients were normalized by the maximum modulus of 
  \(\alpha_{\varphi\varphi}\).
  White stands for negative and black for positive values. The contour 
  lines are \(\pm 0.075,\pm 0.05\), and \(\pm 0.025\)
  for \(\alpha_{\theta\theta}\) and as in Fig. \ref{fig8} for all other 
  coefficients.
  } 
  \label{fig9}
\end{figure} 

A systematic change in the field topology of the mean toroidal field with 
increasing Rossby number can be caused either by a change in the mean
flow or by changes in the mean-field coefficients. The first possibility can 
be ruled out for dipolar models in the low Rossby number regime. For the 
sequence of models displayed in Figs. \ref{fig7} and \ref{fig8}, the mean flow 
does not change significantly. Its influence on the dynamo mechanism is 
relatively weak and not of primary importance 
\citep[see also][]{olson99, schrinner07}. However, the mean-field coefficients,
which are mainly responsible for the generation of the azimuthal field, 
indeed seem to vary in a systematic manner. Figure \ref{fig9} displays the 
dominant dynamo coefficients acting on the toroidal field for model 4, 
model 29, model 31 and model 34. These 
are \(-\gamma_r\) and \(-\gamma_\theta\) describing an advection of the mean 
azimuthal field in radial and latitudinal direction, and \(-\alpha_{rr}\) 
and \(-\alpha_{\theta\theta}\) leading to the generation of toroidal from 
poloidal field.  We normalized all dynamo coefficients by the maximum modulus 
of \(\alpha_{\varphi\varphi}\) for each model. This third diagonal component 
of the \(\alpha\)-tensor is of crucial importance for the generation of 
poloidal from toroidal field and remains almost unchanged for models at 
different \(Ro_\ell\). 

\begin{figure}
  \centering
  \includegraphics{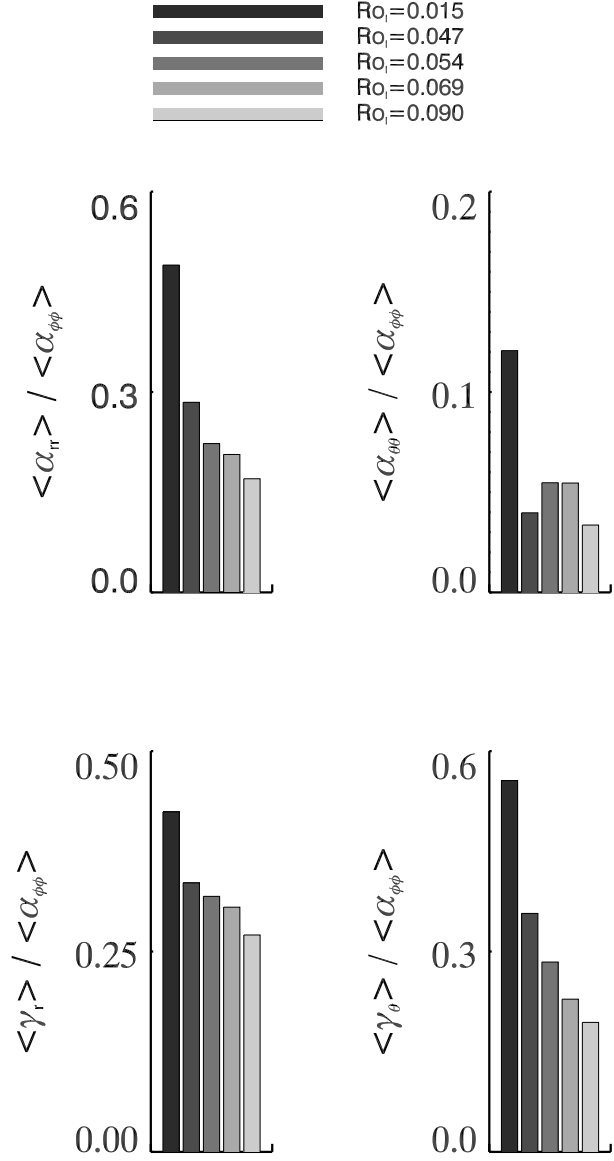}
  \caption{The time and spatially averaged values 
  of \(\alpha_{rr}\), \(\alpha_{\theta\theta}\), \(\gamma_r\), and 
  \(\gamma_\theta\) normalized by the mean value of \(\alpha_{\phi\phi}\) for 
  model 4, model 29, model 31, model 32, and model 34.  
  } 
  \label{fig10}
\end{figure}

The \(\gamma\)--effect advects the mean azimuthal field towards the 
equatorial plane and the outer shell boundary at low latitudes, and in the 
opposite direction at higher latitudes and in deeper layers 
\citep[see also][]{schrinner07}. It is directly related to the columnar 
convection in a spherical shell. A column of fluid elements transported in 
an upwelling towards the outer spherical boundary has to shorten. Because of 
mass conservation, this causes a converging flow towards the equatorial plane. 
Vice versa, the rotational constraint leads to a diverging flow in 
downwellings in deeper layers and at higher latitudes \citep{olson99}. For 
models approaching \(Ro_\ell\approx 0.1\), the rotational constraint is 
relaxed and hence the \(\gamma\)--effect is 
diminished (see also Fig. \ref{fig10}). In particular, the advective velocity 
towards the equatorial plane visualized by the outer contours 
of \(-\gamma_\theta\) in Fig. \ref{fig9} is less prominent for model 34 than 
for model 4. This is consistent with the changes in the topology 
of \(b^0_\varphi\) for these models. 

\begin{figure}
  \centering
  \includegraphics{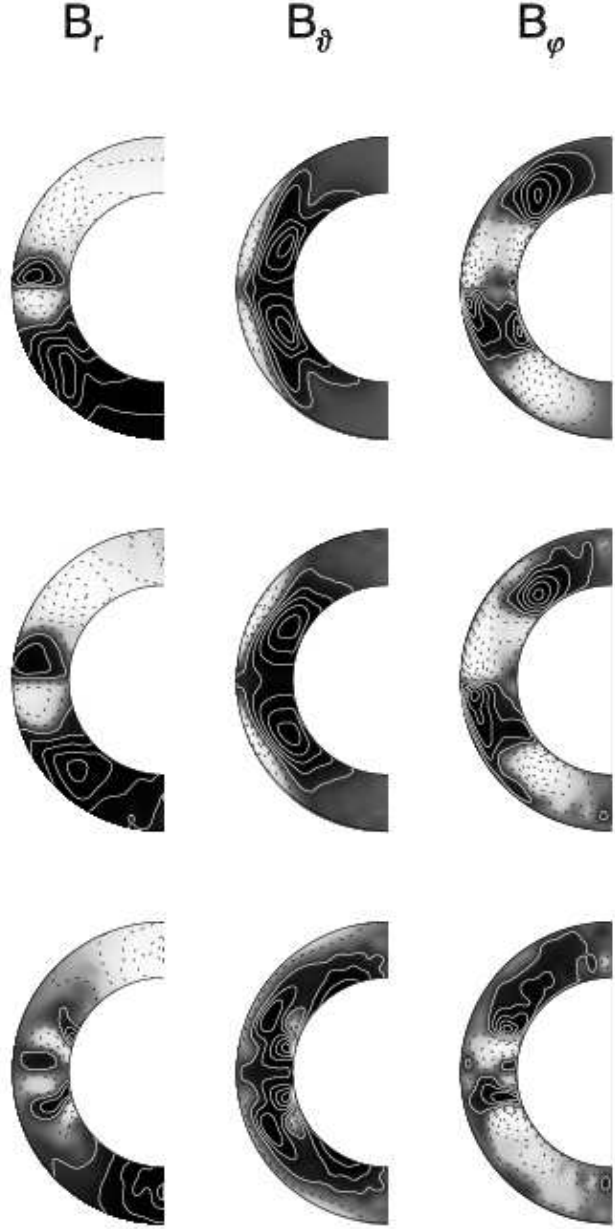}
  \caption{First row: Azimuthally and time-averaged magnetic field for 
  model 34 obtained by direct numerical simulations. Second row: Leading 
  eigenmode derived from (\ref{eq:mf:6}) with the complete dynamo operator for 
  model 34 as defined in (\ref{eq:mf:8}). Third row: Eigenmode derived from 
  (\ref{eq:mf:6}) with a dynamo operator for model 34 in which the 
  \(\gamma\)--effect was artificially suppressed. Each component is normalized 
  by its maximum modulus and the gray-scale coding varies 
  from -1 (white) to +1 (black). The contour lines correspond 
  to \(\pm 0.1, \pm 0.3, \pm0.5, \pm 0.7\), and \(\pm 0.9\).  
  } 
  \label{fig11}
\end{figure}

The significance of the \(\gamma\)--effect in dipolar dynamo models is 
demonstrated by a mean-field calculation, in which \(\mathbf{\gamma}\) was 
arbitrarily suppressed. Figure \ref{fig11} shows the azimuthally and 
time-averaged magnetic field for model 34 obtained from DNS
(first row) and the leading eigenmode of a corresponding 
mean-field calculation based on all mean-field coefficients determined 
(second row). Both are in good agreement. If the dynamo-coefficients related 
to the \(\gamma\)--effect are set to zero in a numerical experiment, the 
result changes substantially. The eigenvalue spectrum is flat and there are two
complex, growing eigenmodes of either symmetry, i.e. the model is no longer 
kinematically stable. The resulting first real, dipolar eigenmode shown in 
the third row of Fig. \ref{fig11} varies on fairly small length scales. 
Moreover, the mean azimuthal flux concentrations at low latitudes near the 
outer boundary characteristic for dipolar dynamo models disappeared.

Similar to the \(\gamma\)-components, \(\alpha_{rr}\) decreases 
considerably with increasing Rossby number, as demonstrated clearly by 
Fig. \ref{fig10}. It is the dominating coefficient responsible for the 
toroidal field generation by an \(\alpha\)--effect. At low Rossby numbers, a 
strong \(\gamma\)--effect leads to the distinctive azimuthal field 
configuration with two flux bundles of inverse polarity close to the 
equatorial plane. The also increased \(\alpha_{rr}\)-component in this dynamo 
regime sustains it against the efficient diffusion due to strong gradients 
necessarily related to this field topology. The 
\(\alpha_{\theta\theta}\)-component, on the other hand,  is on average much 
smaller than \(\alpha_{rr}\) and remains at low level independent of the 
Rossby number.

\clearpage
\subsection{Reason for the bistability and dynamo waves}
\begin{figure}
  \centering
  \includegraphics{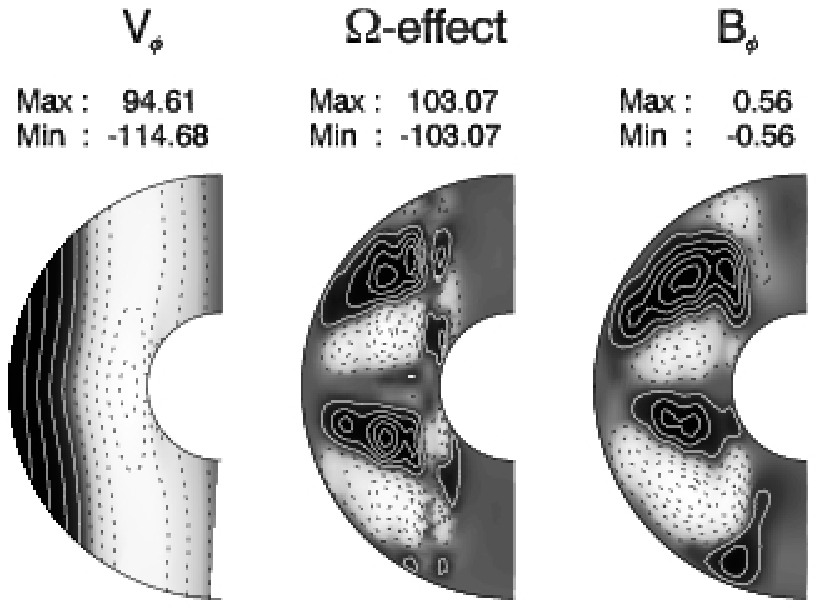}  
  \includegraphics{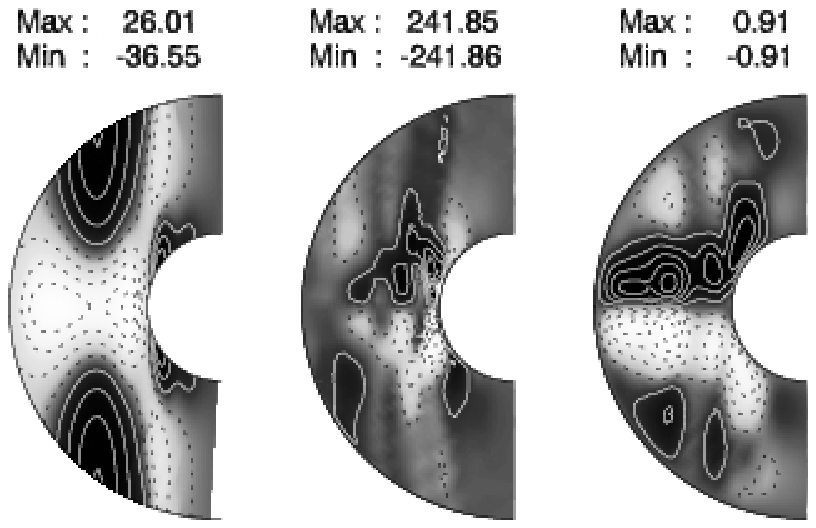}  
\caption{The axisymmetric azimuthal flow, 
the \(\Omega\)--effect expressed as \(r\overline{\vec{B}}_r\,\partial(r^{-1}\overline{\vec{V}}_\phi)/\partial r+r^{-1}\sin\theta\,\overline{\vec{B}}_\theta\,\partial(\sin\theta^{-1}\overline{\vec{V}}_\phi)/\partial\theta\), and the 
axisymmetric azimuthal magnetic field for model 45m (upper panel) and model 
45d (lower panel). The style of the contour plots is explained in the 
caption of Fig. \ref{fig11}.
  } 
  \label{fig12}
\end{figure}

The existence of a dipolar and a multipolar dynamo regime with a sharp 
transition between both at \(Ro_\ell\approx 0.1\) does not depend on the 
choice of the mechanical boundary conditions. The transition here is therefore 
not controlled by the thickness of the Ekman boundary layers 
\citep[as opposed to][]{king09}. However, most of the models with stress-free 
conditions at both boundaries and \(Ro_\ell<0.1\) exhibit a second, 
non-dipolar magnetic-field branch. Stress-free mechanical boundary conditions 
allow for the development of a strong, axisymmetric azimuthal flow, if 
the magnetic field is initially weak. In the presence of stress-free boundary 
conditions, the zonal geostrophic flow can only saturate owing to bulk 
viscosity and magnetic forces. When no-slip boundaries are present, the 
viscous braking of the geostrophic flow occurs mainly
in boundary layers, an effect much stronger (by a factor \(E^{-1/2}\))
than bulk viscous effects  \citep[e.g.][]{morinv06}.
As a result, for stress-free conditions at the inner and 
the outer shell boundaries, even a very weak inertial forcing 
(Reynolds stresses) can yield a very significant geostrophic flow if the 
magnetic field is weak. The zonal flow pattern is then highly geostrophic, 
i.e. \(\overline{V}_\varphi\) is constant on cylinders parallel 
to the rotation axis. Conversely, for the dipolar branch, 
the development of a zonal flow is partly inhibited by Maxwell stresses,
leading to substantial deviations from geostrophy 
\citep[as also observed by][]{busse09}. The difference between both flow 
profiles is visible in Fig. \ref{fig12} for model 45m and model 45d. 
Furthermore, the variation in the mean zonal flow causes differences in 
the \(\Omega\)--effect, 
\(r\overline{\vec{B}}_r\,\partial(r^{-1}\overline{\vec{V}}_\phi)/\partial r+r^{-1}\sin\theta\,\overline{\vec{B}}_\theta\,\partial(\sin\theta^{-1}\overline{\vec{V}}_\phi)/\partial\theta\), and thus in the dynamo mechanism for the toroidal 
field of both branches. For model 45m, the \(\Omega\)--effect correlates 
nicely with the mean azimuthal field (upper panel of Fig. \ref{fig12}), and 
therefore, the model may be characterized as an \(\alpha\Omega\)-dynamo. 
However, for model 45d, the mean azimuthal field is for the most part not the 
result of the \(\Omega\)--effect. In particular the flux portions at larger 
radii and close to the equatorial plane are, if at all, anticorrelated to it. 

The difference in the dynamo mechanism for both branches explains why the 
Rossby number criterion for the dipolarity of the magnetic field applies only 
to the dipolar branch. The Rossby number criterion as formulated in this 
work compares the convective length scale with the Rossby radius.
If the magnetic field is not solely a result of columnar 
convection but its generation mechanism also involves a large-scale zonal 
flow, the Rossby number criterion becomes meaningless. Other examples for 
which it does not apply, presumably for the same reason, can be found 
in \cite{hori2010} and \cite{landeau2011}. 
Similar to models at high \(Ro_\ell\), model 45m lacks the 
particular azimuthal field configuration typical for dipolar or 
`single-mode' dynamos. The magnetic field is governed on this branch by 
several modes and the relative dipole field strength drops 
to \(f_\mathrm{dip}=0.25\). The model is kinematically unstable and the 
magnetic field exhibits quasi-periodic time variations (see Fig. \ref{fig5}).

Dynamo models of this stress-free multipolar branch and those with 
\(Ro_\ell>0.1\) are, for different reasons, not dominated by a dipolar mode. 
The dominance of only one, real dipolar eigenmode associated with the columnar 
flow is broken and in general complex overtones play an essential role in
the dynamics of the magnetic field. If the magnetic Reynolds number is
sufficiently low, the magnetic field evolves in the form of coherent dynamo 
waves. However, with increasing distance of the models from the dynamo 
threshold, the temporal coherence is lost presumably due to the enlarged 
number of relevant modes. A particular dynamo mechanism, on the other hand, 
is not the primary reason for oscillatory dynamos. 
Models 35-38 (\(Ro_\ell>0.1\)) exhibit fairly coherent dynamo waves, but 
they are not of a \(\alpha\Omega\)-type \citep[see][]{schrinner11a}. 
For these examples, oscillatory dynamos are found because the rotational 
constraint is relaxed, i.e. \(Ro_\ell>0.1\), and nevertheless \(Rm\) 
remains moderate (\(Rm\lesssim 200\)). This twofold condition can be 
fulfilled for models with a thin convection zone, for example, and governs 
the transition from steady to oscillatory dynamos already highlighted 
by \cite{goudard08}.

\begin{figure}[ht]
  \centering
  \includegraphics{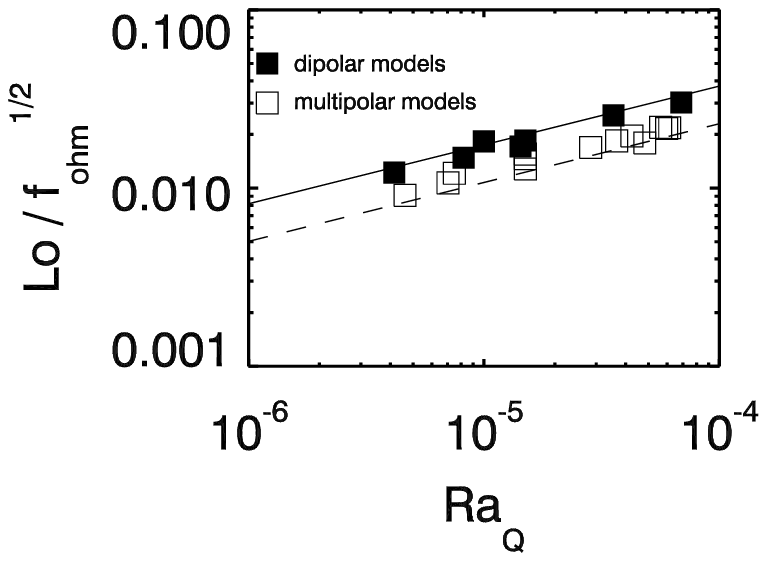}  
  \caption{Lorentz number corrected by the dissipation factor versus the 
  flux based Rayleigh number. The straight lines represent the scaling 
  for dipolar and multipolar models with rigid boundaries
  \citep{christensen10}. 
  Our models with stress-free boundaries (squares stand 
  for dipolar and diamonds for multipolar models)  follow a similar
  scaling. 
  }
  \label{fig14}
\end{figure} 

\cite{christensen06} pointed out that the cube of the magnetic field strength 
for models with rigid boundaries is proportional to the measured buoyancy flux.
The Lorentz number and the flux-based Rayleigh number, 
\(Ra_Q=Ra\,(Nu-1)\,E^2/Pr\), can then be related trough
\begin{equation}
\frac{Lo}{f_\mathrm{ohm}^{1/2}} \propto \,Ra_Q^{1/3},
\label{eq:bi:2}
\end{equation}
where \(f_\mathrm{ohm}\) is the ratio of ohmic to total dissipation.
The coefficient of proportionality was determined, still in the case of rigid 
boundaries, to be \(0.79\) for dipolar models 
and \(0.48\) for multipolar models \citep{christensen10}.
We found, see Fig.\ref{fig14}, that the magnetic field strength for our 
limited sample of stress-free models is consistent with this relation, with 
similar prefactors.

As noted already by \cite{busse11}, there is almost no 
difference in the Nusselt number, and thus in the flux-based Rayleigh number,
for pairs of dipolar and multipolar models in the bistable regime 
(see Table \ref{tab:6}). However, \(f_\mathrm{ohm}\) is 
always smaller for the multipolar branch, i.e. the 
\(\alpha\Omega\)-mechanism is less efficient than the \(\alpha^2\)-mechanism 
related to columnar convection. This deficiency together with the somewhat 
lower prefactor relates to for the lower field strength found for 
these models.

\begin{deluxetable}{crrrrrrrrrrrrrrrr}
\tabletypesize{\scriptsize}
\tablehead{
\multicolumn{17}{c}{Nu and \(f_\mathrm{ohm}\) for models with 
stress-free boundary conditions}
}
\tablecaption{
Nusselt number and ratio of ohmic to total dissipation, 
\(f_\mathrm{ohm}\), for bistable models with stress-free boundary 
conditions \label{tab:6}}
\startdata
Model&43m&43d&45m&45d&46m&46d&47m&47d&48m&48d&49m&49d&51m&51d&52m
&52d\\
Nu&2.0&1.9&2.2&2.4&3.0&3.0&3.0&3.1&2.9&2.9&2.3&2.5&1.3&1.4&4.3&4.2\\
\(f_\mathrm{ohm}\)&0.29&0.48&0.28&0.43&0.32&0.56&0.30&0.56&0.31&0.42&0.25&0.47&0.48&0.58&0.38&0.59
\enddata
\end{deluxetable}

\subsection{Bifurcations between dynamo branches}
\begin{figure}
  \centering 
  \includegraphics[scale=0.3]{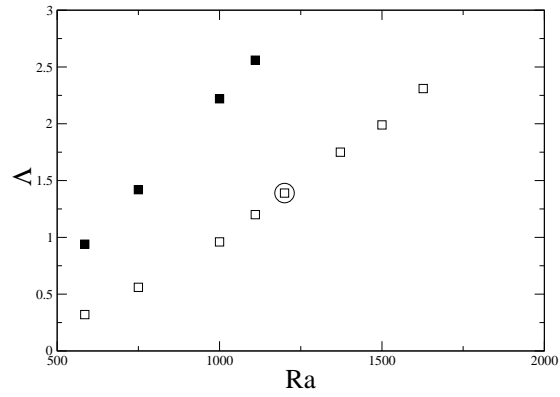}  
  \caption{Evolution of the magnetic field strength, measured by the
    Elsasser number, for both branches as the Rayleigh number is varied at
    fixed Ekman and magnetic Prandtl numbers (\(E=10^{-4}\) and 
    \(Pm=1\)).}
  \label{Ra_Els_branch}
\end{figure}

It is interesting to ponder on the transitions between the 
dipolar and multipolar branch for stress-free models when one 
single control parameter is varied. The two branches are illustrated 
in Fig. \ref{Ra_Els_branch} for a fixed Ekman number of \(E=10^{-4}\)
and magnetic Prandtl number of \(Pm=1\). For both branches, the local Rossby 
number increases with increasing Rayleigh numbers (see Table~1). If the 
Rayleigh number is increased from \(Ra=1110\) (model 52d) on the dipolar 
branch to \(Ra=1200\), the relative dipole field strength collapses 
(the local Rossby number crosses the $Ro_\ell\sim 0.1$ boundary).
The multipolar field configuration then appears to be the only stable 
solution (circle on the figure) and a hysteretic behavior is observed if the 
Rayleigh number is decreased from this state (i.e. model 52m is then obtained).

\begin{figure}
  \centering
  \includegraphics[scale=0.3]{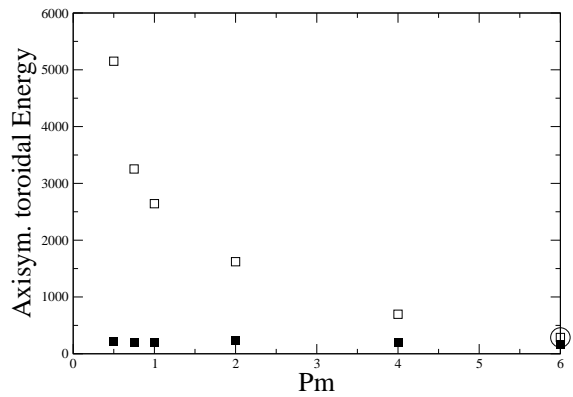}  
  \caption{Axisymmetric toroidal energy density
of the flow as a function of the magnetic Prandtl number at fixed Rayleigh 
number (\(Ra=750\)) and Ekman number (\(E=10^{-4}\)) defined as one half of
the average over the fluid volume of the non-dimensional toroidal velocity
squared.} 
  \label{Vaxis_Pm}
\end{figure}

Interestingly, the transition between both branches is not always as abrupt as 
in Fig. \ref{Ra_Els_branch}. Instead, the two branches can also 
merge more continuously if the zonal geostrophic flow on the 
multipolar branch becomes too weak. This is best demonstrated by varying the 
magnetic Prandtl number keeping all other control parameters fixed. 
Figure \ref{Vaxis_Pm} presents the axisymmetric toroidal kinetic energy 
as a function of the magnetic Prandtl number at fixed Rayleigh 
number (\(Ra=750\)) and Ekman number (\(E=10^{-4}\)). Both branches are 
represented (the multipolar solution corresponding to the larger values). 
If the magnetic Prandtl number is increased the saturated value of the 
geostrophic flow on the multipolar branch decreases. For \(Pm=6\) (circle on 
the figure) the multipolar {solution} is only meta-stable
\citep[see][]{morinv10}. 
It can be observed for a short period of time (enough to assess its amplitude) 
but then switches to the dipolar solution. For this value of the control 
parameter (Pm), only the dipolar solution could be produced.
The saturated amplitude of the geostrophic flow 
is here too small to prevent the growth of the dipolar solution. This 
behavior is typical {for simulations at large} \(Pm\), although the value 
at which the multipolar branch is lost depends on the other
parameters, too.

Finally, besides the transition between both branches, it is worth pondering 
on how these dynamo solutions bifurcate from the purely hydrodynamic solution. 
We have not performed a detailed study of this problem, but 
we observed in Fig. \ref{Vaxis_Pm} that for \(Pm=0.5\), both solutions 
exist, while the hydrodynamic solution is stable (i.e. a small magnetic 
perturbation decays, and a finite amplitude perturbation is needed to 
obtain any of the dynamo solutions). The subcritical bifurcation of the 
dipolar mode at low $Pm$ was already described in \cite{morinv10} 
with no-slip boundary conditions. It is interesting that the system exhibits 
here a triple stability (two dynamo solutions and a purely hydrodynamic mode). 
Obviously the dynamo bifurcation with stress-free boundary conditions 
deserves further studies.

\clearpage
\subsection{Dynamo models versus stellar and planetary dynamos}

It is not known to which extent simplified dynamo models indeed reflect 
physical processes going on in stellar or planetary dynamos. 
All simulations to date are performed in a wrong parameter regime; they are 
therefore not directly comparable with observations. However, scaling 
laws derived from numerical dynamo simulations could help to test their 
reliability. Results obtained from dynamo models may be compared with the 
strength, the geometry and the time dependence of stellar and planetary 
magnetic fields.

The relation between the field strength and the flux-based Rayleigh number
(\ref{eq:bi:2}) proposed by \cite{christensen06} is consistent with the 
field-strength for a class of fast rotating stars and some of the planets 
\citep{christensen09}. However, without further assumptions, it is neither 
applicable to slowly rotating stars, e.g. the Sun, nor to Mercury, Saturn, 
Uranus and Neptune \citep{christensen10}. For theses examples, the field 
strength falls below the predicted value. Moreover, \cite{morin10} observed 
M dwarfs of very similar mass (i.e. energy flux) and rotation rate, but with 
dynamo generated magnetic fields which differ in their field strength and 
their field topologies. \cite{morin11} argued that a different force balance 
could be responsible for the observed bistability, similar to the strong-field 
branch scenario introduced in the context of the geodynamo 
\citep{roberts78,roberts88}. In this study, we point out that differences 
in the dynamo mechanism could likewise lead to two different magnetic field 
branches. The shearing of poloidal field lines (\(\Omega\)--effect) due to a 
strong geostrophic zonal flow may play an important role for the field 
generation in multipolar, but not in dipolar models. A dynamo mechanism based 
on the action of a mean zonal flow leads to models characterized by a more 
variable and a somewhat weaker magnetic field. The bistability investigated in 
this study was found for a wide range of parameter values. It could also 
account for different magnetic fields of stars with similar parameters at a 
Rossby number close to 0.1, which has not been observed yet. 

The Rossby number criterion for the dipolarity of the magnetic field is
consistent with the topology of planetary magnetic fields, except for Uranus 
and Neptune \citep[see][]{christensen10}. Furthermore, it gets some support 
from observations of stellar magnetic fields \citep{morin08}. In particular, a 
decrease of the size of the convection zone may lead to higher local Rossby
numbers and thus to an abrupt change in the field topology, as observed for
early M dwarfs \citep[see Fig. \ref{fig1a} and][]{morin08,donati08,reiners09}. 
Likewise, the multipolar magnetic fields of Uranus and Neptune could be 
compatible with the Rossby number rule, if convection in these 
planets takes place in a thin convection zone, as suggested by 
\cite{stanley04,stanley06}. 

The regime boundary at \(Ro_l\approx 0.1\) separates models with 
a stable dipole field from models exhibiting dipole reversals. Why are  
models of the dipolar regime non-reversing? 
The fact that modes, other than the fundamental mode, are on average
damped does not prevent reversals, which may be triggered by the
coupling with a weakly damped competing dynamo mode  
\citep{petrelis09}. But the `single mode 
property' of these models reported above excludes that different modes 
become critical and explains why a reversal mechanism based on the coupling 
of competing modes has not been identified in these models. 
It has been proposed that the value of \(Ro_\ell\) for the Earth's core may 
be about 0.1 and thus in the vicinity of the regime boundary, which could 
also provide a mechanism for polarity reversals \citep{olson06}.

Whether stellar and planetary magnetic fields are dipolar for the same 
reason as the models could perhaps be assessed studying their time dependence. 
The single mode property should lead to different time scales for the 
variation of the dipolar and the nondipolar field \citep{schrinner11c}, as 
it has been reported for the geomagnetic secular 
variation \citep[e.g.][]{lemouel84}. \cite{tanriverdi11} recently pointed out 
that single mode dynamos may be identified by the spectrum of temporal 
fluctuations of the magnetic energy. If the velocity spectrum is characterized 
by white noise and the evolution of the magnetic field is dominated by only 
one single dynamo mode, a \(\omega^{-2}\) dependence for the spectrum of the 
magnetic field was predicted. This might be verifiable with magnetic field 
data. As discussed above, reversals of the Earth's magnetic field clearly 
indicate that the geodynamo cannot always match the single
mode property.

We found coherent magnetic cycles in our non-dipolar dynamo models only if the 
magnetic Reynolds number is sufficiently low. Similarly, \cite{brown11} reports
oscillations of the magnetic field in an anelastic simulation at fairly 
low magnetic Reynolds number, \(Rm=136\). If  \(Rm\) is increased
in our simulations, the temporal coherence is lost. Moreover, the dynamo 
waves in our models migrate from low latitudes towards the poles 
\citep[see also][]{schrinner11a}, as opposed to solar sunspot regions. 
Given that estimates for the magnetic Reynolds number range from 
\(10^6\) in the photosphere to \(10^{10}\) at the base of the 
convection zone \citep{ossen03}, it remains unclear how the 
temporal coherence visible in the 22-year solar cycle can persist in such a 
highly turbulent environment \citep{jones10}.

\section{Conclusions}

Convection in a rapidly rotating spherical shell is organized in 
quasi-geostrophic columns parallel to the rotation axis. It gives rise to
highly efficient but also very selective dynamo action: only one, real, dipolar
eigenmode of the magnetic field is sustained. This single-mode property
accounts for the dipole dominance and the stability of the dipole field 
in models of this regime. Consequently, the dipolarity of our models 
collapses, if the dominance of the fundamental mode is broken and, in 
general, complex overtones dominate the evolution of the magnetic field. This 
may happen if convection is less constrained by rapid rotation, or, if the 
magnetic field is not solely a result of columnar convection, but a mean zonal 
flow strongly influences the dynamo mechanism. Whether or not differences in 
the topology and the time variability of planetary and stellar magnetic fields 
may be explained by the dichotomy between these two dynamo regimes certainly 
needs to be further explored.

\acknowledgments

We acknowledge discussions with J.F. Donati and J. Morin. MS is grateful for 
financial support from the DFG fellowship SCHR 1299/1-1. LP and ED acknowledge 
financial support from ``Programme National de Physique
Stellaire'' (PNPS) of CNRS/INSU, France. Computations were performed at
CINES and CEMAG computing centers.

\appendix
\section{Appendix}
\label{appendix}
\begin{figure}
  \centering 
  \includegraphics{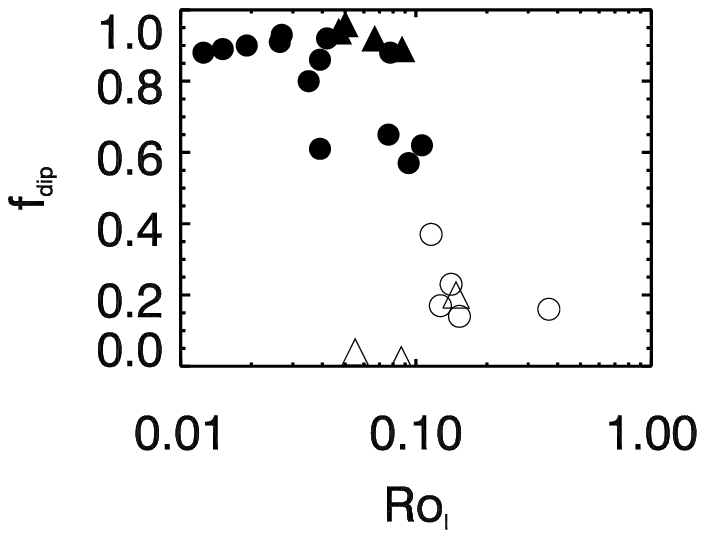}  
  \ \ 
  \includegraphics{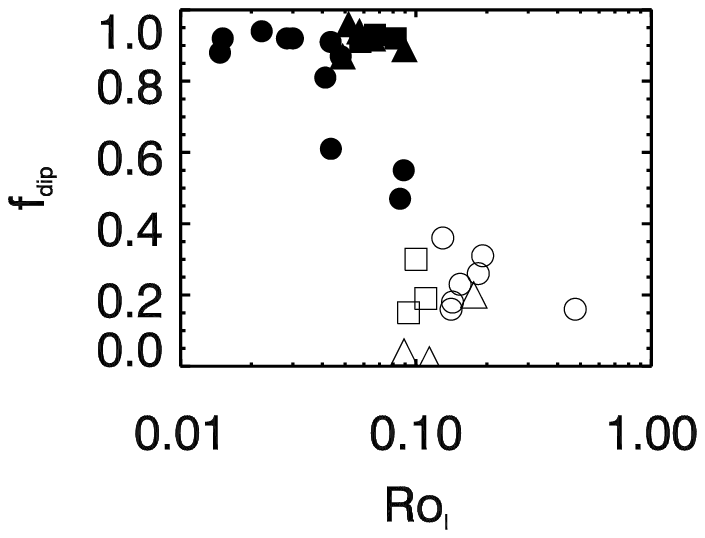}
  \caption{Relative dipole field strength versus local Rossby
    number. On the left, the local Rossby number is based on 
    a length scale derived from the mean harmonic degree of the total flow as 
    introduced by \cite{christensen06}, on the right, 
    it is based on a velocity field and its typical length scale without 
    considering the contribution of the mean toroidal flow component.
    The meaning of the symbols is the same as in Fig. \ref{fig1}.}   
  \label{fig17}
\end{figure}

The definition of the local Rossby number given here relies on a length scale 
derived from the mean harmonic degree of only the poloidal velocity field. 
This intends to filter out the contribution of the mean toroidal flow, which 
is not negligible in models with stress-free mechanical boundary conditions. 
The local Rossby number introduced by \cite{christensen06} based on a length 
scale derived from the mean harmonic degree of the total flow does not allow 
us to distinguish dipolar models from multipolar models equally well 
(Fig. \ref{fig17}, left panel). 

In this study we argued that the typical length scale of convection relative 
to the Rossby radius strongly influences the topology of the magnetic field 
in our models. The definition of a local Rossby number based on a velocity 
field, in which only the axisymmetric toroidal contribution is canceled out 
is maybe more appropriate to test this argument than the definition used 
throughout in the text. The right panel of Fig. \ref{fig17} shows that a 
local Rossby number defined in this way seems to serve equally well to 
distinguish dipolar from multipolar models.   
\clearpage  
\bibliographystyle{apj}
\bibliography{schrinner}
\end{document}